\documentclass[11pt]{article}
\usepackage{amsmath}
\normalbaselines

\renewcommand{\Box}{\mbox{}}
\newtheorem{thm}{Theorem}[section]
\newtheorem{lem}{Lemma}[section]

\begin{document}
\bibliographystyle{unsrt}

\def\bea*{\begin{eqnarray*}}
\def\eea*{\end{eqnarray*}}
\def\ba{\begin{array}}
\def\ea{\end{array}}
\count1=1
\def\be{\ifnum \count1=0 $$ \else \begin{equation}\fi}
\def\ee{\ifnum\count1=0 $$ \else \end{equation}\fi}
\def\ele(#1){\ifnum\count1=0 \eqno({\bf #1}) $$ \else \label{#1}\end{equation}\fi}
\def\req(#1){\ifnum\count1=0 {\bf #1}\else \ref{#1}\fi}
\def\bea(#1){\ifnum \count1=0   $$ \begin{array}{#1}
\else \begin{equation} \begin{array}{#1} \fi}
\def\eea{\ifnum \count1=0 \end{array} $$
\else  \end{array}\end{equation}\fi}
\def\elea(#1){\ifnum \count1=0 \end{array}\label{#1}\eqno({\bf #1}) $$
\else\end{array}\label{#1}\end{equation}\fi}
\def\cit(#1){
\ifnum\count1=0 {\bf #1} \cite{#1} \else 
\cite{#1}\fi}
\def\bibit(#1){\ifnum\count1=0 \bibitem{#1} [#1    ] \else \bibitem{#1}\fi}
\def\ds{\displaystyle}
\def\hb{\hfill\break}
\def\comment#1{\hb {***** {\em #1} *****}\hb }

\newcommand{\TZ}{\hbox{\bf T}}
\newcommand{\MZ}{\hbox{\bf M}}
\newcommand{\ZZ}{\hbox{\bf Z}}
\newcommand{\NZ}{\hbox{\bf N}}
\newcommand{\RZ}{\hbox{\bf R}}
\newcommand{\CZ}{\,\hbox{\bf C}}
\newcommand{\PZ}{\hbox{\bf P}}
\newcommand{\QZ}{\hbox{\rm eight}}
\newcommand{\HZ}{\hbox{\bf H}}
\newcommand{\EZ}{\hbox{\bf E}}
\newcommand{\GZ}{\,\hbox{\bf G}}

\font\germ=eufm10
\def\goth#1{\hbox{\germ #1}}
\vbox{\vspace{38mm}}

\begin{center}
{\LARGE \bf The $Q$-operator and Functional Relations 
of the Eight-vertex Model at Root-of-unity $\eta = \frac{2m K}{N}$ for odd $N$ 
}\\[10 mm] 
Shi-shyr Roan \\
{\it Institute of Mathematics \\
Academia Sinica \\  Taipei , Taiwan \\
(email: maroan@gate.sinica.edu.tw ) } \\[25mm]
\end{center}
\begin{abstract}
Following Baxter's method of producing $Q_{72}$-operator, we construct the $Q$-operator of the root-of-unity eight-vertex model for the crossing parameter $\eta = \frac{2m K}{N}$ with odd $N$ where $Q_{72}$ does not exist. We use this new $Q$-operator to study the functional relations in the Fabricius-McCoy comparison between the root-of-unity eight-vertex model and the superintegrable $N$-state chiral Potts model. By the compatibility 
of the constructed $Q$-operator with the structure of Baxter's eight-vertex (solid-on-solid) SOS model, we verify the set of functional relations of the root-of-unity eight-vertex model using the explicit form of the $Q$-operator and fusion weights of SOS model.  
\end{abstract}
\par \vspace{5mm} \noindent
{\it 1999 PACS}:  05.50.+q, 02.30.Dk, 75.10Jm \par \noindent

\setcounter{section}{0}
\section{Introduction \label{sec.In}}
\setcounter{equation}{0}
It is known that the (zero-field) eight-vertex model was explicitly solved by Baxter \cite{B71, B72} by the method of $TQ$-relation. Here the eight-vertex model is assumed with periodic boundary condition and {\it even} chain-size $L$, (this restriction applies  throughout this paper unless otherwise stated). In fact there are  many $Q$-operators in this context \cite{Bax}, and the first discovered one by Baxter in 1972 \cite{B72} (valid for both even and odd chain-site $L$), denoted by $Q_{72}$, was on the special ''root of unity'' case 
$$
2N \eta_{72} = 2m_1 K + {\rm i} m_2 K' , \ ~ \ ~ \ ~ {\rm i} : = \sqrt{-1}
$$
where $K, K'$ are the complete elliptic integrals, $N, m_1, m_2$ are integers, and  $\eta_{72}$ is the (crossing) parameter of the eight-vertex model. Using the $Q_{72}$-operator for $m_2=0$, Fabricius and McCoy computed the degeneracy of eight-vertex eigenvalues, a property previously found in \cite{De01a, De01b}, then proposed the functional equations for the eight-vertex model at $\eta_{72}$ in \cite{FM02, FM04} as an analogy with the set of functional equations known in the $N$-state chiral Potts model (CPM) \cite{BBP}. The Fabricius-McCoy comparison between CPM and the eight-vertex model at $\eta_{72}$ was further analysed about their common mathematical structures in \cite{R04, R05o}, where the effort led to the discovery of Onsager-algebra symmetry of superintegrable CPM. However, by employing the $Q_{72}$-operator in the study of the special "root-of-unity" eight-vertex model, the conjectural functional relations, strongly supported by computational evidences, are valid only in the cases for either even $N$  or both $N$, $m$ odd in the above $\eta_{72}$-expression (\cite{FM02} (3.10), \cite{FM04}(3.1), \cite{FM41} (4.20)). Those cases do not include some ''other'' important root-of-unity type of  eight-vertex model that appeared in a sequence of 1973 papers \cite{B73I, B73II, B73III} by Baxter in the study of the eight-vertex eigenvectors, where the parameter $\eta$ satisfies the following ''{\it root of unity}'' condition (\cite{B73I} (9), \cite{B73II} (6.8), \cite{B73III} (1.9) or \cite{B01} (113))\footnote{The letters $N, L, m$ in this paper are the $L, N, m_1$ in \cite{B73I, B73II, B73III}. Here for simplicity, we consider only the case $m_2 =0$  in \cite{B73I, B73II, B73III} by easier calculations of Jacobi theta functions, (indeed no essential difficulties could arise for other cases by using the modified elliptic functions in \cite{B73I} (10)).}:
\be
 \eta = \frac{2 m K}{N} , \ ~ \ ~ \ ~ N \ {\rm and} \ m = {\rm odd} , \ ~ \ ~ {\rm gcd}(N, m ) = 1 .
\ele(eta)
In the present work, we study the eight-vertex models with the parameter $\eta$ restricted only in the above case (\req(eta)), which for convenience, will be loosely called the root-of-unity eight-vertex model throughout this paper.  As noticed in \cite{FM02}, the $Q_{72}$-operator and the $Q$-operator in \cite{B73I, B73II, B73III} are different; in fact, it was shown in section II of \cite{FM02} that $Q_{72}$-operator does not exist when  $\eta$ satisfies (\req(eta)). Thus, the quest for a proper $Q$-operator in accordance with  "symmetry" of the eight-vertex model for the root-of-unity $\eta$ in (\req(eta)) appears to be a compelling problem for its solution. 

The purpose of this paper is to construct a $Q$-operator of the eight-vertex model for the parameter $\eta$ in (\req(eta)), and to conduct the functional-relation study of the eight-vertex model as a parallel theory of the CPM. In the present paper we provide a mathematical justification about the conjectural functional relations of the root-of-unity eight-vertex model by the explicit $Q$-operator constructed along the line, but not the same, as the $Q_{72}$-operator in \cite{B72}. Indeed, we produce the $Q$-operator by following the same mechanism in \cite{R06Q} of constructing the $Q$-operator of root-of-unity six-vertex  at the $N$th root-of-unity anisotropic parameter $q$  with odd $N$, which can be regarded as the vanishing elliptic nome limit of (\req(eta)). Consequently, they share some remarkable qualitative and semi-quantitative resemblances in the functional-relation study of root-of-unity symmetry of the theory. Furthermore, our $Q$-operator (more precisely, the $Q_R$-operator) coincides with the eight-vertex $Q$-operator recently found by Fabricius in \cite{F06}, but with the different specified values for the free parameter, subsequently a subtle difference occurs in the expression of $Q$-functional equation  about the related involution  (see formula (3) and Section 3 in \cite{F06}, and (\req(Qeq)) (\req(Cdef)) of this paper). One special character of the $Q$-operator in this work is that it possesses essential features appeared in Baxter's 1973 papers  \cite{B73I, B73II, B73III}, where he invented the original techniques to convert the eight-vertex model to an ice-type solid-on-solid (SOS) model, and derived the equation of eigenvectors by a generalized Bethe ansatz method.
It is worth noting that the three original papers by Baxter in 1973 subsequently laid the foundation to many exceptional developments in the theory of quantum integrable systems, among which were the restricted SOS-model \cite{ABF}, algebraic Bethe ansatz of the eight-vertex model \cite{TakF}, the theory of elliptic quantum group \cite{FVn, FVc}, and the recently developed analytic theory of functional relations in the eight-vertex/SOS-model in \cite{BM}. As the main observation of this work, we find a $Q$-operator of the root-of-unity eight-vertex model built upon the 
Baxter's eigenvectors in \cite{B73II} with parameters $s, t$ taking certain special values so that results in \cite{B73I, B73II, B73III} can be employed in calculations when verifying the relation between the $N$th-fusion operator and $Q$-operator, which serves as the $Q$-operator-constraint about the ''root-of-unity'' symmetry of the theory. 
In this way, by assuming (as to be the case by numerical evidences for small $L$) the non-singular property of a certain $M(v_0)$-matrix (see (\req(Qv0)) in the paper), the whole set of functional relations can be successfully justified, much as seen mathematically in \cite{R06Q} for a similar discussion of the root-of-unity six-vertex model. As a consequence, the conjectural $Q$-functional relation raised in \cite{FM02, FM04} is verified for the root-of-unity eight-vertex model with $\eta$ in (\req(eta)), but using a different involution appeared in the formula.

This paper is organized as follows. In section \ref{sec.8v}, we recall known results and conjectures in the root-of-unity eight-vertex model. We first briefly review some basic facts of the eight-vertex model in subsection \ref{ssec.qd8v}. In subsection \ref{ssec.Fus}, we construct the fusion matrices from the fused $L$-operators, and derive the fusion relations of the eight-vertex model. In the root-of-unity cases, we discuss the relationship between $TQ$-, $QQ$- and $Q$-functional relations, in particular the equivalent relation between the $QQ$- and  $Q$-functional relations. 
Section \ref{sec.8vFR} contains the main results of this paper about the $Q$-operator and the functional relations in the root-of-unity eight-vertex model, as a parallel theory to the superintegrable CPM and root-of-unity six-vertex model established in \cite{R05o, R06Q}. 
We first briefly summarize some main features in the eight-vertex SOS-model and fusion weights in \cite{B73I, B73II, B73III, DJKMOp, DJKMO}, needed for later discussions. 
In subsection \ref{ssec.8Q}, by imitating Baxter's method of producing  $Q_{72}$-operator in \cite{B72}, we derive another $Q$-operator, different from  $Q_{72}$, of the eight-vertex model with the parameter $\eta$ only in (\req(eta)). Using results known in the eight-vertex SOS-model, we then in subsection \ref{ssec.pfQQ}  
show the validity of functional relations using the constructed $Q$-operator.   We close in section \ref{sec.F} with some concluding remarks.

\section{Eight-vertex Model and the Fusion Operators \label{sec.8v} }
We start with some basic facts about the eight-vertex model in subsection \ref{ssec.qd8v}. Then in subsection \ref{ssec.Fus}, we construct the fusion matrix, and establish the fusion relations  of the eight-vertex model; for the root-of-unity eight-vertex model, we discuss the relationship between $TQ$-, $QQ$- and $Q$-functional relations.

\setcounter{equation}{0}

\subsection{Formalism and quantum determinant of the eight-vertex model \label{ssec.qd8v}}
First we review some basic notions in the eight-vertex model, (for more details, see any standard reference listed in the biography, such as \cite{Bax, TakF} and references therein). This also serves to establish the notation.

The Boltzmann weights of the eight-vertex model are described by the homogeneous coordinates of an elliptic curve in the complex projective 3-space $\PZ^3$:
\be
E ~ (= E_{\triangle, \gamma}) : ~ ~ a^2 + b^2 - c^2 - d^2= 2 \triangle (ab + cd) , \ \ cd = \gamma ab , \ \ \ \ \  a:b:c:d \in \PZ^3
\ele(Cur)
where $\triangle, \gamma \in \CZ$. One can parameterize the above elliptic curve in terms of Jacobi theta functions of moduli $k, k'= (1-k^2)^{1/2}$ with the complete elliptic integrals $K, K'$:
$$
\begin{array}{lll}
a  & = \Theta (2 \eta)  \Theta (v - \eta) H (v + \eta) &= \rho(v)   {\rm sn}(v+ \eta) , \\
b  & = \Theta (2 \eta)  H (v - \eta) \Theta (v + \eta) &= \rho(v) {\rm sn}(v- \eta) , \\
c  & = H (2 \eta)  \Theta (v - \eta) \Theta (v + \eta) &= \rho(v) {\rm sn}(2 \eta) , \\
d  & = H (2 \eta)  H (v - \eta) H (v + \eta) &= \rho(v)  k {\rm sn}(2 \eta){\rm sn}(v- \eta) {\rm sn}(v+ \eta) , 
\end{array}
$$
where $H ( v ) = \vartheta_1 (\frac{v}{ 2K} , \tau ), \Theta ( v ) = \vartheta_4 (\frac{v}{ 2K} , \tau )$, $\rho(v) = k^{\frac{1}{2}} \Theta (2 \eta)  \Theta (v - \eta) \Theta (v + \eta) $ with $\tau = \frac{{\rm i}K'}{K}$, and the relations 
\be
H( v - 2K ) = H(- v) = - H(v), \ ~ \ \Theta( v - 2K ) = \Theta (- v) = \Theta (v).
\ele(HT)
Hence $H ( v + 2N \eta ) = H (v), \Theta ( v + 2N \eta ) = \Theta (v)$ for $\eta$ in (\req(eta)). 
The parameters $\triangle, \gamma$ in (\req(Cur)) are given by $\triangle = \frac{{\rm cn} (2 \eta) {\rm dn} (2 \eta)}{1+ k {\rm sn}^2 (2 \eta)},  \gamma = k {\rm sn}^2 (2 \eta)$,  ((43), (44) and (54) in \cite{B73I}). The uniformizing variable $v \in \CZ$ is called the spectral parameter. The elliptic curve $E$ in (\req(Cur)) is biregular to the complex torus of $\CZ$ quotiented by the lattice $4K \ZZ + 2K' \ZZ$:
$E =  \CZ/ (4K \ZZ + 2K' \ZZ)$. The quotient of $E$ by the $\ZZ_2$-automorphism, $v \mapsto v - 2K$, (corresponding to $a: b: c: d \mapsto -a: -b : c : d$), is the torus $\CZ/(2K \ZZ + 2K' \ZZ)$.
In this paper, we shall consider the order-$N$ automorphism $U$ of elliptic curve $E$ with $\eta$ in (\req(eta)), and the elliptic function $h(v)$ : 
\bea(l) 
h(v) = \Theta (0)\Theta (v) H(v) , \ \ \ U: v \mapsto  v - 2 \eta . 
\elea(hU)

Using the elliptic coordinates $a,b,c,d$ in (\req(Cur)), one defines  the $L$-operator of the eight-vertex model, which is the matrix of $\CZ^2$-auxiliary and $\CZ^2$-quantum space 
\be
L (v)  =  \left( \begin{array}{cc}
        L_{0,0}  & L_{0,1}  \\
        L_{1,0} & L_{1,1} 
\end{array} \right) (v) , \ \ v \in \CZ  , 
\ele(8VL)
where entries $L_{i,j}$ are the $\CZ^2$-(quantum-space) operators
$$
\begin{array}{l}
L_{0,0} = \left( \begin{array}{cc}
        a  & 0\\
        0& b 
\end{array} \right), \ L_{0,1} = \left( \begin{array}{cc}
        0  & d\\
        c& 0
\end{array} \right) , \ L_{1,0} = \left( \begin{array}{cc}
        0  & c\\
        d& 0 
\end{array} \right) , \ L_{1,1} = \left( \begin{array}{cc}
        b  & 0\\
        0& a 
\end{array} \right). 
\end{array}
$$
Equivalently to say, the eight-vertex weights are
\bea(ll)
R(+, + | +, + ) = R(-, - | -, - ) = a , & R(+, + | -, - ) = R(-, - | +, + )  = b , \\
R(+, - | -, + ) = R(-, + | +, - ) = c , & R(+, - | +, - ) = R(-, + | -, + )  = d . 
\elea(8wt)
The $L$-operator (\req(8VL)) satisfies the YB relation,
\be
R_{\rm 8v}(v'-v) (L(v') \bigotimes_{aux}1) ( 1
\bigotimes_{aux} L(v)) = (1
\bigotimes_{aux} L(v))( L(v')
\bigotimes_{aux} 1) R_{\rm 8v}(v'-v),
\ele(8YB)
with the $R$-matrix 
$$
R_{\rm 8v} (v) = \left( \begin{array}{cccc}
          a(v+\eta) & 0 & 0 & d (v+\eta) \\
          0 & b (v+\eta) & c (v+\eta) & 0 \\
         0 & c (v+\eta) & b (v+\eta) & 0 \\
          d (v+\eta) & 0 & 0 & a (v+\eta)
         \end{array}   \right) 
$$
(see, e.g. \cite{TakF}). Then the monodromy matrix of chain-size $L$,
\be
M (v) = L_1 (v) \otimes  \cdots \otimes   L_L (v) =  \left( \begin{array}{cc} A (v)  & B (v) \\
      C (v) & D (v)
\end{array} \right),
\ele(8M)
again satisfies the YB equation (\req(8YB)). The traces of monodromy matrices 
\be
T(v) := {\rm tr}_{\rm aux} M (v) = A ( v) + D (v), ~ \ ~ v \in \CZ, 
\ele(8T)
form a commuting  family of $\stackrel{L}{\otimes} \CZ^2$-operators, called the transfer matrix of the eight-vertex model, which commutes with the spin-reflection operator $R$ and the $S$-operator:
\be
~ [T(v) , S ] = [T(v) , R ] = 0 ,  \ ~ \ ~ \ ~ \ {\rm where} \ \ S = \prod_{\ell=1}^L \sigma^{z}_\ell \ , ~ \ ~ \ R = \prod_{\ell=1}^L \sigma^{x}_\ell \ .
\ele(TRS)
As the $R$-matrix $R_{\rm 8v}$ at $v = - 2 \eta$ is of rank-one:
$$
R_{\rm 8v} (- 2\eta) = - {\rm sn} ( 2 \eta) \left( \begin{array}{cccc}
          0 & 0 & 0 & 0 \\
          0 & 1 & -1 & 0 \\
         0 & -1 & 1 & 0 \\
          0 & 0 & 0 & 0
         \end{array}   \right),
$$
the quantum determinant of (\req(8M)) is defined by (\req(8YB)) with $v' = v - 2 \eta$:
$$
\begin{array}{ll}
{\rm det}_q M (v) \cdot R_{\rm 8v}(- 2\eta) &= R_{\rm 8v}(- 2\eta) (M (v- 2\eta) \bigotimes_{aux}1) ( 1
\bigotimes_{aux} M(v )) \\
&= (1 \bigotimes_{aux} M(v ))(M_L(v - 2\eta)
\bigotimes_{aux} 1) R_{\rm 8v}(- 2\eta) .
\end{array}
$$
The above relation is equivalent to the following set of relations:
\bea(rll)
&B  ( v) A (v- 2 \eta) = A  ( v) B (v- 2 \eta), & D  ( v) C (v- 2 \eta) = C  ( v) D (v- 2 \eta) , \\
&A (v- 2 \eta ) C (v ) = C (v - 2 \eta) A (v ), & B (v- 2 \eta) D (v ) = D (v- 2 \eta) B (v  ), \\
{\rm det}_q M(v) 
=& D (v)A (v- 2 \eta) - C (v )B (v- 2 \eta) &= A (v)D (v- 2 \eta) - B (v )C (v- 2 \eta) \\
=& A (v- 2 \eta) D (v ) - C (v- 2 \eta) B (v)&= D (v- 2 \eta) A (v ) - B (v- 2 \eta) C (v );
\elea(qdet8)
with the explicit form of quantum determinant: ${\rm det}_q M(v) = h (v + \eta)^L h (v- 3 \eta)^L$.

\subsection{ Fusion relation of the eight-vertex model \label{ssec.Fus}}
As in the six-vertex model \cite{R06Q}, we now construct the eight-vertex fusion matrices $T^{(J)} (v)$ for a non-negative integer $J$ from the $J$th fused $L$-operator $L^{(J)}(v)$, which is a matrix of $\CZ^2$-quantum and $\CZ^J$-auxiliary space defined as follows.
Denote the standard basis $|\pm 1 \rangle$ of the $\CZ^2$-auxiliary space of $L(v)$ in (\req(8VL)) by $\widehat{x}= |1 \rangle, \widehat{y}= |-1 \rangle$, and its dual basis  by $ x, y $. 
For non-negative integers $m$ and $n$, $\widehat{x}^m \widehat{y}^n$ is the completely symmetric $(m+n)$-tensor of $\CZ^2$ defined by
$$
{m+n \choose n} \widehat{x}^m \widehat{y}^n = \underbrace{\widehat{x}\otimes \ldots \otimes \widehat{x}}_{m} \otimes \underbrace{\widehat{y}\otimes \ldots \otimes \widehat{y}}_{n} + \ {\rm all \ other \ terms \ by \ permutations} , 
$$
similarly for $x^m y^n$. For $J \geq 1$, the $\CZ^J$-auxiliary space is the space of completely symmetric $(J-1)$-tensors of $\CZ^2$ with the canonical basis $e^{(J)}_k$ and the dual basis $e^{(J) *}_k$:
\be
e^{(J)}_k = \widehat{x}^{J-1-k} \widehat{y}^k, \ \ e^{(J) *}_k = {J-1-k \choose k} x^{J-1-k} y^k , \ \ k=0, \ldots, J-1.
\ele(Cjb)
By the first and third relations in (\req(qdet8)) for $L=1$, or equivalently,
$$
\begin{array}{ccl}
\langle x^2 | L(v) \otimes_{aux} L (v-2 \eta ) | \widehat{x} \wedge \widehat{y} \rangle &= \langle y^2 | L(v) \otimes_{aux} L (v-2 \eta ) | \widehat{x} \wedge \widehat{y} \rangle &= 0 , \\
\langle x \otimes y | L(v) \otimes_{aux} L (v-2 \eta ) | \widehat{x} \wedge \widehat{y} \rangle &= \langle y \otimes x | L(v) \otimes_{aux} L (v-2 \eta ) | - \widehat{x} \wedge \widehat{y} \rangle &= \frac{1}{2} {\rm det}_q L(v) ,
\end{array}
$$
where $\widehat{x} \wedge \widehat{y} = \frac{1}{2} ( \widehat{x} \otimes \widehat{y} - \widehat{y} \otimes \widehat{x})$, the following relations hold: 
$$
\langle e^{(3)*}_k | L(v) \otimes_{aux} L (v-2 \eta ) | \widehat{x} \otimes \widehat{y} \rangle= \langle e^{(3)*}_k | L(v) \otimes_{aux} L (v-2 \eta ) | \widehat{y} \otimes \widehat{x} \rangle  \ \ {\rm for} \ k=0,1,2 .
$$
As a consequence for an integer $J \geq 2$, and $v_i = \widehat{x}$ or $\widehat{y}$ for $1 \leq i \leq J-1$, we have 
\bea(ll)
&\langle e^{(J)*}_k | L(v) \otimes_{aux} \cdots \otimes_{aux} L(v -2(J-3)\eta) \otimes_{aux} L(v -2(J-2)\eta)) | v_1 \otimes \cdots \otimes v_{J-1} \rangle \\
= & \langle e^{(J)*}_k | L(v) \otimes_{aux} \cdots \otimes_{aux} L(v -2(J-3)\eta) \otimes_{aux} L(v -2(J-2)\eta)) | v_{\sigma_1}  \otimes \cdots \otimes v_{\sigma_{J-1}} \rangle
\elea(gdet)
where $0 \leq k \leq J-1$, and $\sigma$ is an arbitrary permutation of $\{1, \ldots, J-1 \}$. 
By using the basis (\req(Cjb)) of the $\CZ^J$-auxiliary space, the fused $L^{(J)}$-operator of eight-vertex model,
$L^{(J)}(v) = \bigg( L^{(J)}_{k, l} (v)\bigg)_{0 \leq k, l \leq J-1}$, is defined by the following $\CZ^2$-(quantum-space)-operators $L^{(J)}_{k, l}(v)$:
\be
L^{(J)}_{k, l} (v) = \frac{\langle e^{(J)*}_k | L(v) \otimes_{aux} L(v- 2\eta)\otimes_{aux} \cdots \otimes_{aux} L(v - 2(J-2) \eta ) | e^{(J)}_l \rangle}{ \prod_{i =0}^{J-3} h( v - (2i +1) \eta ) } .
\ele(Lj)
For $J=3$, by additive formulas of theta functions (\cite{Bax} (15.4.25) (15.4.26)),
$$
\begin{array}{ll}
\Theta (u) \Theta (v) \Theta (a-u) \Theta (a-v) - H(u) H(v) H(a-u) H(a-v) = \Theta (0) \Theta (a) \Theta (u-v) \Theta (a-u-v) , \\
H (v) H (a-v) \Theta (u) \Theta (a-u) - \Theta (v) \Theta (a-v) H(u) H(a-u) = \Theta (0) \Theta (a) H (v-u) H (a-u-v) ,
\end{array}
$$
$\langle e^{(3)*}_k | L(v) \otimes_{aux} L (v-2 \eta ) | e^{(3)}_l \rangle$ for $0 \leq k, l \leq 2$ are all divisible by $h(v - \eta)$. Indeed 
$L^{(3)}_{k, l} ( =L^{(3)}_{k, l} (v))$ are expressed by
$$
\begin{array}{ll}
L^{(3)}_{0, 0} =  \frac{\Theta(2 \eta)^2}{\Theta(0)} \left( \begin{array}{cc}
        H(v+\eta) \Theta(v-3\eta) & 0\\
        0& \Theta(v+\eta) H(v-3\eta)
\end{array} \right) , & L^{(3)}_{0, 0} \leftrightarrow L^{(3)}_{2, 2} , \ \Theta (v+j\eta) \leftrightarrow H(v+j \eta) , \\
L^{(3)}_{1, 0} 
= H (4\eta) \left( \begin{array}{cc}
        0&\Theta (v - \eta)^2  \\
       H(v -\eta)^2 & 0
\end{array} \right),& L^{(3)}_{1, 0} \leftrightarrow L^{(3)}_{1, 2} , \ \Theta (v+j\eta) \leftrightarrow H(v+j \eta) , \\
L^{(3)}_{2, 0} =  \frac{H^2 (2 \eta)}{\Theta(0)}  
\left( \begin{array}{cc}
        \Theta(v+\eta) H(v-3\eta)& 0\\
        0& H(v+\eta) \Theta (v-3\eta)
\end{array} \right),& L^{(3)}_{2, 0} \leftrightarrow L^{(3)}_{0, 2} , \ \Theta (v+j\eta) \leftrightarrow H(v+j \eta) , \\
L^{(3)}_{0, 1} =  \frac{\Theta (2\eta)H(2\eta)}{\Theta(0)} \left( \begin{array}{cc}
        0 & H(v+\eta) H(v-3\eta)   \\
       \Theta(v+\eta)\Theta(v-3\eta)& 0
\end{array} \right) , 
&  L^{(3)}_{0, 1} \leftrightarrow L^{(3)}_{2, 1} , \ \Theta (v+j\eta) \leftrightarrow H(v+j \eta) , \\
L^{(3)}_{1, 1}  = \Theta(4\eta)  \left( \begin{array}{cc}
         H(v-\eta)\Theta(v-\eta)& 0\\
        0& \Theta(v-\eta)H(v-\eta)
\end{array} \right). \\
\end{array}
$$
By this,  all $L^{(J)}_{k, l} (v)$ in (\req(gdet)) are elliptic functions without poles (\cite{DJKMO} Lemma 2.3.1). Using $L^{(J)}(v)$ as the local operator, one defines the $J$th fusion matrix $T^{(J)}(v)$ as the trace of the monodromy matrix:
\be
T^{(J)} (v) = {\rm tr}_{\CZ^J} (\bigotimes_{\ell=1}^L  L^{(J)}_\ell ( v)), \ \ L^{(J)}_\ell ( v) = L^{(J)} ( v) \ {\rm at \ site} \ \ell ,
\ele(tauj)
which form a family of commuting operators of the quantum space $\stackrel{L}{\otimes} \CZ^2$ with $T^{(2)} (v) = T (v)$. One can derive 
the recursive fusion relation among $T^{(J)}$'s as the case of six-vertex model (see, e.g. Sect. 3 of \cite{R06F}). Regard the auxiliary-space tensor $\CZ^2 \otimes \CZ^J$ as a subspace of $\stackrel{J+1}{\otimes} \CZ^2$, the auxiliary space $\CZ^{J+1}$ as a subspace of $\CZ^2 \otimes \CZ^J$  by identifying the basis elements: 
$$
e^{(J+1)}_{k+1} = \frac{1}{{J \choose k+1}} \bigg( {J-1 \choose k+1} \widehat{x} \otimes e^{(J)}_{k+1}  +
{J-1 \choose k} \widehat{y} \otimes e^{(J)}_k  \bigg) , \ \ k=-1, \ldots, J-1.  
$$ 
Denote $
f^{(J-1)}_k := \widehat{x} \otimes e^{(J)}_{k+1}  -  \widehat{y} \otimes e^{(J)}_k $ for $ 0 \leq k \leq J-2$.
Then $e^{(J+1)}_l, f^{(J-1)}_k$ form a basis of $\CZ^2 \otimes \CZ^J$, with the dual basis $e^{(J+1)*}_l, f^{(J-1)*}_k$ expressed by
$$
\begin{array}{ll}
e^{(J+1)*}_{k+1} = x \otimes e^{(J)*}_{k+1}  +  y \otimes e^{(J)*}_k , & f^{(J-1)*}_k = \frac{1}{{J \choose k+1}} \bigg( {J-1 \choose k} x \otimes e^{(J)*}_{k+1} - \ {J-1 \choose k+1}y \otimes e^{(J)*}_k   \bigg).
\end{array}
$$
One has 
$$
\begin{array}{l}
L^{(J+1)}_{k, l}(v) = \langle e^{(J+1)*}_k | L^{(J+1)} (v) | e^{(J+1)}_l \rangle = \frac{1}{h(v- (2J-3) \eta) } \langle e^{(J+1)*}_k | L^{(J)}(v) \otimes_{aux} L(v-2(J-1)\eta ) | e^{(J+1)}_l \rangle ; \\
\langle e^{(J+1)*}_l | L^{(J)}(v) \otimes_{aux} L(v-2(J-1)\eta ) | f^{(J-1)}_k \rangle = 0 ; \\
\langle f^{(J-1)*}_k | L^{(J)}(v) \otimes_{aux} L(v-2(J-1) \eta ) | f^{(J-1)}_l \rangle = h(v-(2J-1) \eta ) \langle e^{(J-1)*}_k | L^{(J-1)}(v)  | e^{(J-1)}_l \rangle .
\end{array}
$$
Then follows the recursive fusion relation by setting $T^{(0)}=0$, $T^{(1)}= h(v + \eta)^L$:
\be
T^{(J)}(v) T^{(2)}(v-2(J-1)\eta ) = h^L(v-(2J-1) \eta ) T^{(J-1)}(v) + h^L(v- (2J-3) \eta) T^{(J+1)} (v), \ \ J \geq 1.
\ele(8fu)
Since the chain-size $L$ is even, by (\req(HT)) one finds the periodic property of $T ^{(J)}$:
$$
T(v - 2K)= T(v) , \ ~ \ ~ T ^{(J)}(v - 2K ) = T ^{(J)}(v). 
$$

The eigenvalues of $T(v)$ are computed in \cite{B71, B72, Bax} using an auxiliary $Q$-matrix, i.e.,a commuting family, $Q(v)$ for $v \in \CZ$, with $[T(v), Q(v)]=0$, and the Baxter's $TQ$-relation 
$$
T (v) Q (v) = h^L(v- \eta) Q( v+2 \eta) + h^L( v+ \eta) Q(v - 2 \eta) ,
$$
equivalently, 
\bea(l)
T (v) Q (v) = \widetilde{h} (v)^L Q( U^{-1} v ) + \widetilde{h}  ( U^{-1} v )^L Q(U v) , \ \ \widetilde{h} (v):= h(v-\eta) ,
\elea(8TQ)
where $U$ is defined in (\req(hU)). 
By (\req(8fu)), (\req(8TQ)) and using an induction argument, one finds the $T^{(J)}Q$-relation for $ J \geq 0$:
\bea(l)
T^{(J)}(v) = Q( U^{-1} v )Q( U^{J-1}v )\sum_{k=0}^{J-1} \bigg( \widetilde{h} (U^{k-1} v)^L Q(U^{k-1}v)^{-1} Q(U^k v)^{-1}  \bigg).
\elea(8TjQ)
In the root-of-unity case (\req(eta)), the relation (\req(8TjQ)) in turn yields the boundary fusion relation:
\be 
T^{(N+1)}(v) =  T^{(N-1)}(U v) + 2  \widetilde{h} (U^{-1} v)^L .
\ele(8bFu)
Furthermore, it is expected that some $Q$-operator will encode essential features about the root-of-symmetry of the eight-vertex model, much as in the study of Onsager-algebra symmetry of superintegrable $N$-state CPM in \cite{R05o}. Parallel to the functional equation of the chiral Potts transfer matrix (\cite{BBP} (4.40)), the root-of-unity $Q$-operator of  eight-vertex model conjecturally satisfies the $Q$-functional equation (\cite{FM02} (3.10), \cite{FM04}(3.1)):  
\be
Q ( C v ) = M_0 Q ( v ) \sum_{k=0}^{N-1} \bigg( \widetilde{h} (U^k v)^L Q (U^k v)^{-1} Q (U^{k+1} v)^{-1}  \bigg)
\ele(Qeq)
where $C$ is an order-$2$ automorphism of elliptic curve $E$ commuting with $U$, and $M_0$ is some normalized matrix independent of $v$. By (\req(8TjQ)), the $Q$-functional equation (\req(Qeq)) is the same as the $N$th $QQ$-relation ( \cite{FM04} (3.11), \cite{R05o} (44), \cite{R06Q} Theorem 3.1)
\be
 T^{(N)}(U v) = M^{-1}_0 Q ( C v ) Q(v)  , 
\ele(QQN)
which is equivalent to either one of the following $QQ$-relations:  
\be
T^{(J)}(U v) + T^{(N-J)}(U^{J+1} v) = M_0^{-1} Q ( C U^J v )Q(v) , ~ \ ~ \ ~ 0 \leq J \leq N .
\ele(QQ)
In the next section, we are going to construct a $Q$-operator satisfying (\req(QQN)) with the elliptic automorphism $C$ defined by\footnote{Note that the automorphism $C$ here differs from the $C_{72} : v \mapsto v - {\rm i} K'$ for the $Q_{72}$-operator in \cite{FM04}. Hence the $Q$-operator in this paper carries a different nature from the Baxter's $Q_{72}$ in \cite{B72}.} 
\be
C : v \mapsto  v- 2K .
\ele(Cdef)
{\bf Remark}. In the vanishing elliptic nome limit, $K \rightarrow \frac{\pi}{2},  K' \rightarrow \infty$, and the $\eta$ in (\req(eta)) tends to the $N$th root-of-unity $q$  for odd $N$ in the six-vertex model with the spectral parameter $s$ as the limiting value of  $-e^{\frac{-\pi {\rm i} v }{2K}}$. Then the automorphism (\req(Cdef)) corresponds the $s$-involution, $C_0 (s)= -s$, in the XXZ limit. The $Q$-operator of the six-vertex model at a such $q$, satisfying functional relations corresponding to (\req(Qeq))-(\req(QQ)) with the involution $C_0$, was constructed in \cite{R06Q} Theorem 4.2.

\section{The $Q$-operator and Verification of Functional Relations of the Eight-vertex Model at $\eta = \frac{2 m K}{N}$ \label{sec.8vFR} }
\setcounter{equation}{0}
In this section, we discuss the functional relations of the eight-vertex model incorporated with the root-of-unity property. First we construct in subsection \ref{ssec.8Q} the $Q$-operator of the eight-vertex model for the root-of-unity $\eta$ (\req(eta)) following the Baxter's method of producing $Q_{72}$ in \cite{B72}. The $Q$-operator obtained here  differs from $Q_{72}$, but will accord with the root-of-unity symmetry of the eight-vertex model in the sense that the set of functional relations is valid for this special $Q$-operator. 
Furthermore, the structure of the $Q$-operator can be fed into the scheme of Baxter's eight-vertex SOS model in the eight-vertex-eigenvector discussion  in \cite{B73I, B73II, B73III}. 
In subsection \ref{ssec.pfQQ}, we give a mathematical verification about the conjectured functional relations. The methods rely on results in \cite{B73I, B73II, B73III} about the equivalence between the eight-vertex model and SOS-model, plus the study of fusion weights of the eight-vertex SOS-model in \cite{DJKMOp, DJKMO}, which we now briefly summarize as follows.

For the convenience, we introduce the vectors and covectors as in \cite{B73II} (6.4):
\be
| v \rangle = \left( \begin{array}{c}
        H(v)\\
        \Theta (v)
\end{array} \right) , \ \ \langle v | = \bigg( \Theta (v), - H (v) \bigg), \ \ \ v \in \CZ .
\ele(|v)
For $s \in \CZ$ and integers $l \in \ZZ$, we define the local vector in \cite{B73II} (C.13):
\be
\Phi_{l, l+ \mu } (= \Phi_{l, l+ \mu } (v)) = | s + 2l \eta + \mu (\eta - v) \rangle , \ ~ \   \mu = \pm 1 , 
\ele(Phi)
and the product-vector (\cite{B73II} (3.2)) for a set of integers $l_1, \ldots, l_{L+1}$ with  $l_{\ell+1} - l_{\ell} = \pm 1$ for $ 1 \leq \ell \leq L$ :
\be
\psi (l_1, \ldots, l_{L+1})(v) = \Phi_{l_1, l_2} \otimes \Phi_{l_2, l_3} \otimes \cdots \otimes \Phi_{l_L, l_{L+1} }.
\ele(psi)

In the study of eigenvectors of the eight-vertex transfer matrices, Baxter converted it to a SOS model \cite{B73I, B73II, B73III} so that the eight-vertex weights $R(\alpha , \beta | \lambda, \mu)$ (\req(8wt)) are changed to the SOS "lattice-weights" $W(m, m' |l, l')$ by employing the local vectors $\Phi_{m, l}$ in (\req(Phi)) through the relation
\be
\sum_{\beta, \mu } R( \alpha , \beta | \lambda, \mu)(v) \Phi_{l, l'}(v')_\beta z_{m', l'}(v, v')_\mu = \sum_{m}  W(m, m' |l, l')(v) \Phi_{m, m'}(v')_\alpha  z_{m, l}(v, v')_\lambda ,
\ele(RW)
where $\alpha , \beta, \lambda, \mu = \pm 1$ , $l, l', m, m' \in \ZZ$ with $|l-l'|=|m-m'|=|m-l|=|m'-l'|=1$, and $z_{m, l}$ are the vectors\footnote{The $z_{l-1, l}$ here is in \cite{B73II} (B.26) (C.13), which differs from \cite{B73II} (6.5) by a factor. }
$$
z_{l+1, l}(v, v') = | s + v-v'  + 2l \eta  \rangle , \ \ z_{l-1, l}(v, v') = | s -v+v' + 2l \eta  \rangle .
$$ 
Indeed, the Boltzmann weights $W(m, m' |l, l')$ for {\it integers} $m, m', l', l$ are zeros except $\lambda  = m-l$, $ \mu = m'-l'$, $\alpha = m' -m$, $\beta = l'-l$ equal to $\pm 1$;
and the non-zero weights are
\bea(cll)
W(l \pm 1, l \pm 2 | l , l \pm 1 )(v) &= \frac{h (v + \eta )}{h ( 2 \eta) }, & (\lambda = \mu = \alpha = \beta =  \pm 1);  \\
W(l \pm 1, l  | l, l \mp 1) (v) &=  \frac{h(s-K + 2 (l \mp 1) \eta) h ( v - \eta ) }{h ( 2 \eta) h(s-K + 2 l \eta)} ,  & ( \lambda = \mu = -\alpha = -\beta  = \pm 1 ) ; \\ 
W(l \pm 1, l  | l, l \pm 1) (v) &= \frac{h ( s- K + 2 l \eta  \mp (v- \eta))}{h ( s- K + 2 l \eta )}, & (\lambda = -\mu = -\alpha  = \beta = \pm 1 ) ,
\elea(wSOS)
((C. 14) and (C30)-(C.33) of \cite{B73II}, or (2.1.4a)- (2.1.4c) of \cite{DJKMO}\footnote{The $W_{11}(a, b, c, d)$, $u, \lambda, \xi$ in \cite{DJKMO} are equal to $W (a, b | d, c)$, $\frac{v - \eta}{2 \eta}, 2\eta , \frac{s -K}{2 \eta}$ here, respectively.} ).  In the case (\req(eta)), by using (\req(RW)), the eight-vertex transfer matrix (\req(8T)) transfers the vector $\psi ( l_1, \ldots , l_{L+1} )$ in (\req(psi)) to a linear combination of product-vectors:
\be
T(v)\psi ( l_1, \ldots , l_{L+1} )(v') = \sum_{m_\ell} \{\prod_{\ell=1}^L W(m_\ell, m_{\ell+1}|l_\ell, l_{\ell+1})(v) \} \psi (m_1, \ldots, m_{L+1})(v'),
\ele(TSOS) 
where the summation runs over integers $m_\ell$'s with $m_{\ell +1}= m_\ell \pm 1$, $m_\ell = l_\ell \pm 1$ for all $\ell$, and $l_{L+1} - l_1= m_{L+1}-m_1  \equiv 0 \pmod{N}$ ((1.5), (1.9) and (1.11) of \cite{B73III}).

For a positive integer $J$, one can derive the $J$th-fusion weights, $W^{(J)}(m, m' | l, l')(v)$,  of the SOS model such that $W^{(2)}(m, m' | l, l')(v) = W(m, m' | l, l')(v)$ in (\req(wSOS)). Indeed, $W^{(J)}(m, m' | l, l')(v)$ are zeros except $ \lambda^{(J)}, \mu^{(J)} = J-1 - 2k$ for $0 \leq k \leq J-1$, and $ |\alpha| = |\beta| =1$, where $\lambda^{(J)} = m-l$, $\mu^{(J)} = m'-l'$,  $\lambda = m' -m$, $\mu = l'-l$:  
$$
\put (-35 , 0 ){\line(1, 0){70}}
\put (0 , -35 ){\line( 0, 1){70}}
\put ( -2, -44 ){\shortstack{ $\alpha$ }}
\put ( -2, 40 ){\shortstack{ $\beta$ }}
\put (-55 , -2 ){\shortstack{ $ \lambda^{(J)} $ }}
\put (43 , -2){\shortstack{ $\mu^{(J)}$ }}
\put (12 , 12){\shortstack{ $l'$ }}
\put (-18 , 12){\shortstack{ $l$ }}
\put (12 , -18){\shortstack{ $m'$ }}
\put (-18 , -18){\shortstack{ $m$ }}
$$
By formulas (2.1.16), (2.1.20) in \cite{DJKMO}\footnote{The $W^{(J)}(m, m' | l, l')(v)$ here is equal to $W_{1, J-1}(m, m', l',l | u)$ in \cite{DJKMO}.} , the non-zero $J$th-fusion weights  are given by  
\bea(rll)
W^{(J)}(l-1, l | l'-1, l')(v)= &  \frac{ h(s-K +(l+l'+J-1) \eta )h(v+(l-l'-J+2) \eta ) }{h(2 \eta)h(s-K + 2 l \eta)} &  \\
=& \frac{ h(s-K +(2l- \mu^{(J)} +J-1) \eta )h(v+(\mu^{(J)}-J+2) \eta ) }{h(2 \eta)h(s-K + 2 l \eta)} , & ( \lambda^{(J)} =  \mu^{(J)},  \alpha =  \beta = 1 ) ;   \\
W^{(J)}(l+1, l | l'+1 , l')(v) =&  \frac{h(s-K + (l+l'- J+1) \eta) h(v+(l'-l-J+2) \eta) }{h(2 \eta) h(s-K + 2 l \eta) }  \\
=& \frac{h(s-K + (2l- \mu^{(J)}- J+1) \eta) h(v- (\mu^{(J)}+J-2) \eta) }{h(2 \eta) h(s-K + 2 l \eta) } , & (\lambda^{(J)} = \mu^{(J)}, \alpha= \beta = -1) ; \\
W^{(J)}(l+1, l | l'-1, l')(v) = &\frac{h((l'-l+J-1) \eta)h(s-K +(l+l'+J-2) \eta -v  )}{h(2 \eta) h(s-K + 2 l \eta)}  \\
 =& \frac{h((-\mu^{(J)}+J-1) \eta)h(s-K +(2l-\mu^{(J)}+J-2) \eta -v  )}{h(2 \eta) h(s-K + 2 l \eta)} , & (\lambda^{(J)} = \mu^{(J)}+2 , \alpha=   - \beta = -1) ; \\
W^{(J)}(l-1, l | l'+1, l')(v)  =& \frac{h((l-l'+J-1) \eta) h(s-K +(l+l'-J+2) \eta + v  ) }{h(2 \eta)h(s-K + 2 l \eta)}  \\
=&  \frac{h((\mu^{(J)}+J-1) \eta) h(s-K +(2l-\mu^{(J)}-J+2) \eta + v  ) }{h(2 \eta)h(s-K + 2 l \eta)} , & (\lambda^{(J)} = \mu^{(J)}-2 , \alpha=   - \beta = 1 ).
\elea(WJ)
For the $N$th "root-of-unity" $\eta$ in (\req(eta)), as in (\req(TSOS)), the $J$th fusion matrix $T^{(J)} $ in (\req(tauj)) is related to weights $W^{(J)}(m, m'|l, l')$ through  product-vectors in (\req(psi)) by 
\be
T^{(J)} (v) \psi (l_1, \ldots , l_{L+1})(v') 
= \sum_{m_\ell} \{\prod_{\ell=1}^L W^{(J)}(m_\ell, m_{\ell+1}|l_\ell, l_{\ell+1})(v) \} \psi (m_1,  \ldots, m_{L+1})(v'),
\ele(TSOS)
the summation being integers $m_\ell$'s with $m_{\ell +1}= m_\ell \pm 1$, $m_\ell - l_\ell = J-1 -2k_\ell$, $0 \leq k_\ell \leq J-1$, for all $\ell$, and $l_{L+1} - l_1= m_{L+1}-m_1  \equiv 0 \pmod{N}$ (\cite{DJKMOp}, \cite{DJKMO} Theorem 2.3.3 ). Later in this paper, we shall work on $T^{(N)}$ with the variable evaluating at $v- \eta$, then the corresponding SOS-weights in (\req(WJ)) become  
\bea(l)
W^{(N)}(l-1, l | l'-1, l')(v- 2 \eta )   
= \frac{ h(-K +(2l- \mu^{(N)} -1) \eta )h(v+\mu^{(N)} \eta ) }{h(2 \eta)h(-K + 2 l \eta)} , (\lambda^{(N)} = \mu^{(N)} ,  \alpha= \beta = 1) ; \\
W^{(N)}(l+1, l | l'+1 , l')(v- 2 \eta)  
= \frac{h(-K + (2l- \mu^{(N)}+1) \eta) h(v-\mu^{(N)} \eta) }{h(2 \eta) h(-K + 2 l \eta) } , (\lambda^{(N)} = \mu^{(N)}, \alpha= \beta = -1); \\
W^{(N)}(l+1, l | l'-1, l')(v-2 \eta )  
 = \frac{h((-\mu^{(N)}-1) \eta)h(-K +(2l-\mu^{(N)}) \eta -v  )}{h(2 \eta) h(-K + 2 l \eta)} , (\lambda^{(N)} = \mu^{(N)}+2 , \alpha= - \beta = -1) ; \\
W^{(N)}(l-1, l | l'+1, l')(v- 2 \eta)  
 = \frac{h((\mu^{(N)}-1) \eta) h(-K +(2l-\mu^{(N)}) \eta + v  ) }{h(2 \eta)h(-K + 2 l \eta)} , (\lambda^{(N)} = \mu^{(N)}-2 , \alpha= - \beta = 1 ). \\ 
\elea(WN)
Here we use the relation $h( v + N \eta ) = - h(v)$ by the condition (\req(eta)) on $\eta$.

\subsection{ The $Q$-operator of the eight-vertex model for $\eta = \frac{2 m K}{N}$
\label{ssec.8Q}}

As the construction of $Q_{72}$ in \cite{B72}, we start with the ${\sf S }$-, $\widehat{\sf S }$-operator of $\CZ^N$-auxiliary and $\CZ^2$-quantum space, ${\sf S }= ({\sf S }_{i,j})_{i, j \in \ZZ_N} $, $\widehat{\sf S }= (\widehat{\sf S }_{i,j})_{i, j \in \ZZ_N} $, where $\ZZ_N = \ZZ/N\ZZ$, and ${\sf S }_{i,j}$, $\widehat{\sf S }_{i,j}$ are $\CZ^2$-operators. Here the $\CZ^N$-basis of the auxiliary space are indexed by $\ZZ_N$.
The general forms of ${\sf Q}_R, {\sf Q}_L$-matrices are 
\be
{\sf Q}_R= {\rm tr}_{\CZ^N} ( \bigotimes_{\ell =1}^L {\sf S}_{\ell}), \ ~ {\sf Q}_R= {\rm tr}_{\CZ^N} ( \bigotimes_{\ell =1}^L \widehat{\sf S}_{\ell} \ , ~ \ ~ {\rm where} ~ \ {\sf S}_{\ell}, \widehat{\sf S}_{\ell}= {\sf S}, \widehat{\sf S} \ {\rm at ~ site} \ \ell,
\ele(QRL)
with $T{\sf Q}_R = {\rm tr}_{\CZ^2 \otimes \CZ^N} ( \bigotimes_{\ell =1}^L {\sf U}_{\ell})$ , ${\sf Q}_L T= {\rm tr}_{\CZ^2 \otimes \CZ^N} ( \bigotimes_{\ell =1}^L \widehat{\sf U}_{\ell})$, where ${\sf U}_{\ell}, \widehat{\sf U}_{\ell}= {\sf U}, \widehat{\sf U}$ at site $\ell$, and ${\sf U}, \widehat{\sf U}$ are the matrices  of $\CZ^2 \otimes \CZ^N $-auxiliary and $\CZ^2$-quantum space
$$
{\sf U} = \left( \begin{array}{cc}
        L_{0,0} {\sf S } & L_{0,1} {\sf S} \\
        L_{1,0} {\sf S} & L_{1,1} {\sf S }
\end{array} \right) , \ \ \widehat{\sf U} = \left( \begin{array}{cc}
         \widehat{\sf S } L_{0,0} &  \widehat{\sf S}L_{0,1} \\
        \widehat{\sf S} L_{1,0}  & \widehat{\sf S }L_{1,1} 
\end{array} \right) .
$$
The operator $T{\sf Q}_R $, ${\sf Q}_L T $ will decompose into the sum of two matrices if we can find a $2N$ by $2N$ scalar matrix ${\sf M}$ (independent of $v$)
such that 
\be
{\sf M}^{-1} {\sf U} {\sf M} = \left( \begin{array}{cc}
        {\sf A}  & 0  \\ 
         {\sf C} & {\sf D}
\end{array} \right), \ \ {\sf M}^{-1} \widehat{\sf U} {\sf M} = \left( \begin{array}{cc}
        \widehat{\sf A}  & 0  \\ 
         \widehat{\sf C} & \widehat{\sf D}
\end{array} \right) .
\ele(MUM)
The above required form is unaffected by postmultiplying ${\sf M}$ by a upper blocktriangular matrix. Together with a similar transformation of ${\sf S}$, we can in general choose
\be
{\sf M} = \left( \begin{array}{cc}
        I_N  & \delta \\
        0 & I_N 
\end{array} \right) , \ \ \delta = {\rm dia} [\delta_0, \cdots, \delta_{N-1}] .
\ele(Mdef)
Hence
$$
{\sf M}^{-1} {\sf U} {\sf M} = \left( \begin{array}{cc}
L_{0,0} {\sf S } - \delta L_{1, 0} {\sf S }, & L_{0,0} {\sf S } \delta  - \delta L_{1,0} {\sf S } \delta  +  L_{0,1} {\sf S }
       - \delta  L_{1,1} {\sf S}   \\ 
       L_{1,0} {\sf S} , &   L_{1,0} {\sf S } \delta  + L_{1,1} {\sf S }
\end{array} \right) .
$$
The condition for non-zero  ${\sf S}_{i,j}$'s in above with vanishing upper blocktriangular matrix  is (\cite{B72} (C.10)):
$$
\left( \begin{array}{cc}
        a   \delta_j  -\delta_i b ,  &  d - c \delta_i \delta_j   \\
        c - d \delta_i \delta_j & b \delta_j -   \delta_i a
\end{array} \right) {\sf S}_{i, j} = 0 , \ \ i, j \in \ZZ_N ,
$$
which in turn yields 
\be
(a^2 + b^2 - c^2 - d^2) \delta_i \delta_j = ab (\delta^2_i + \delta^2_j) - cd (1+ \delta^2_i \delta^2_j) .
\ele(dij)
If we set $\delta_i = k^{\frac{1}{2}} {\rm sn} (u)$, by (\req(Cur)), then follows  $\delta_j = k^{\frac{1}{2}} {\rm sn} (u \pm 2 \eta )$. For 
$\delta_i = k^{\frac{1}{2}} {\rm sn} (u)$, $\delta_j = k^{\frac{1}{2}} {\rm sn} (u \pm 2 \eta )$,  using the general formulae
$$
\begin{array}{ll}
{\rm sn} A  {\rm sn} B -  {\rm sn} C   {\rm sn} D  = \frac{\Theta (0) \Theta (A+B) H (A-D) H(B-D) }{k \Theta (A) \Theta (B) \Theta  (C) \Theta  (D) } , &
1 - k^2 {\rm sn} A  {\rm sn} B   {\rm sn} C  {\rm sn} D  = \frac{\Theta (0) \Theta (A+B) \Theta (A-D) \Theta (B-D) }{ \Theta (A) \Theta (B) \Theta  (C) \Theta  (D) } ,
\end{array}
$$
when $A+B = C+D$, one can derive 
$$
{\sf S}_{i,j} (= {\sf S}_{i,j} (v) ) = | u \pm (\eta - v) \rangle \tau_{i, j} , \ ~ ~ \ ~ \tau_{i, j} = (\tau^1_{i, j}, \tau^2_{i, j}) ,
$$
and the relations 
$$
\begin{array}{l}
(L_{0,0} {\sf S }_{i, j})(v) - \delta_i ( L_{1, 0} {\sf S }_{i, j})(v) = h (v - \eta)  \frac{\Theta (u \pm 2 \eta)  }{    \Theta  (u)  } {\sf S }_{i, j}(v + 2 \eta ) , \\
\delta_j ( L_{1, 0} {\sf S }_{i, j})(v) + (L_{1,1} {\sf S }_{i, j})(v) = 
 h (v + \eta)   \frac{ \Theta (u)  }{ \Theta (u \pm 2 \eta)  } {\sf S }_{i, j}(v-2 \eta) , 
\end{array}
$$
where $h(v)$ is in (\req(hU)). We choose $s$ so that $\delta_i  = k^{\frac{1}{2}} {\rm sn}(s + 2i \eta ), i \in \ZZ_N$, are $N$-distinct diagonal entries of $\delta$ in (\req(Mdef)). Then\footnote{The ${\sf S}_{i,j}$ of (\req(SsR)) here is equal to the $(\widehat{S}_R)_{k,l}$ of (22) in \cite{F06}, where $k, l, L$ and the parameter $t$ correspond to $i+1, j+1, N$ and $s- \eta$ in this paper. However in the discussion of $Q_L$-operator, there is a slight difference about the construction. Indeed, the $\widehat{\sf S}_{i,j}$ of (\req(LSs)) in this paper differs from the $(\widehat{S}_L)_{k,l}$ in \cite{F06} (28) by the multiplication of $\left( \begin{array}{cc} 0 & -1 \\ 1 & 0 \end{array} \right)$, $\widehat{\sf S}_{i,j}= (\widehat{S}_L)_{k,l}\left( \begin{array}{cc} 0 & -1 \\ 1 & 0 \end{array} \right)$, with the identification of parameters: $\widehat{s}= t + \eta$.} 
\be
{\sf S}_{i,j} (v) = \left\{\begin{array}{ll}
 | s + 2 i \eta + (j - i ) (\eta - v) \rangle \tau_{i, j}  &{\rm if} \ j- i= \pm 1 , \\
0 & {\rm otherwise} . \end{array}  \right.  
\ele(SsR)
with the arbitrary parameters $\tau_{i, j}$.  The ${\sf A}, {\sf D}$ in (\req(MUM)) are related to ${\sf S}$ by 
$$
{\sf A} (v) = h(v - 2 \eta ) {\sf d}^{-1} {\sf S} (v + 2 \eta ) {\sf d} , \ ~  
{\sf D} (v) = h(v + 2 \eta ) {\sf d} {\sf S} (v - 2 \eta ) {\sf d}^{-1} ; 
$$
where ${\sf d}$ is the diagonal matrix ${\rm dia}. [ \Theta (s), \Theta (s + 2\eta), \cdots , \Theta ( s + 2(N-1) \eta ]$. This imply  
$$
T(v) {\sf Q}_R (v) = h(v - 2 \eta )^L  {\sf Q}_R ( v + 2 \eta ) + h(v + 2 \eta )^L {\sf Q}_R ( v - 2 \eta ) .
$$

With the similar discussion for ${\sf M}^{-1} \widehat{\sf U} {\sf M}$, the non-zero $\widehat{\sf S}_{i,j}$ again yields the relation (\req(dij)). For 
$\delta_i = k^{\frac{1}{2}} {\rm sn} (u)$, $\delta_j = k^{\frac{1}{2}} {\rm sn} (u \pm 2 \eta )$, one arrives the expression  
$$
\widehat{\sf S}_{i, j} ( = \widehat{\sf S}_{i, j} (v))  = \widehat{\tau}_{i, j} \langle  u  \pm (\eta + v)| , \ ~ ~ \ ~ \widehat{\tau}_{i, j} = (\widehat{\tau}_{i, j ; 1}, \widehat{\tau}_{i, j ; 2})^t , 
$$
and the relations
$$
\begin{array}{l}
(\widehat{\sf S }_{i, j} L_{0,0} )(v) - \delta_i ( \widehat{\sf S }_{i, j} L_{1, 0} )(v) =  h (v + \eta )  \frac{ \Theta (u \pm 2\eta ) }{ \Theta (u)  } \widehat{\sf S }_{i, j} (v- 2\eta) 
 , \\
\delta_j ( \widehat{\sf S }_{i, j} L_{1, 0} )(v) + (\widehat{\sf S }_{i, j} L_{1,1} )(v) = 
h(v-\eta)\frac{ \Theta (u)   }{ \Theta (u \pm 2 \eta)  } \widehat{\sf S }_{i, j} (v + 2 \eta) .
\end{array}
$$
We define the $\delta$ in (\req(Mdef)) by $N$-distinct numbers $\delta_i  = k^{\frac{1}{2}} {\rm sn}(\widehat{s} + 2i \eta ), i \in \ZZ_N$, for some $\widehat{s}$. Then $\widehat{\sf S}_{i,j}$ are given by   
\be
\widehat{\sf S}_{i,j} (v) = \left\{\begin{array}{ll}
 \widehat{\tau}_{i, j} \langle  \widehat{s} + 2i \eta  + (j-i)  (\eta + v)  |   &{\rm if} \ j- i= \pm 1 , \\
0 & {\rm otherwise} . \end{array}  \right.  
\ele(LSs)
and related to $\widehat{\sf A}, \widehat{\sf D}$ in (\req(MUM)) by 
$$
\begin{array}{ll}
\widehat{\sf A} (v) = h(v + 2 \eta ) \widehat{\sf d}^{-1} \widehat{\sf S} (v - 2 \eta ) \widehat{\sf d} , &\widehat{\sf D} (v) = h(v - 2 \eta ) \widehat{\sf d} \widehat{\sf S} (v + 2 \eta ) \widehat{\sf d}^{-1} ,
\end{array}
$$
where $\widehat{\sf d}= {\rm dia}. [ \Theta (\widehat{s}), \Theta (\widehat{s} + 2\eta), \cdots , \Theta (\widehat{s} + 2(N-1) \eta ]$.
Hence 
$$
{\sf Q}_L (v) T(v) = h(v - 2 \eta )^L  {\sf Q}_L ( v + 2 \eta ) + h(v + 2 \eta )^L {\sf Q}_L ( v - 2 \eta ) .
$$

To construct a $Q(v)$ matrix from the ${\sf Q}_R$- and ${\sf Q}_L$-operator, as in \cite{B72}(C28) it suffices to find $\widehat{s}, s$ in (\req(LSs)), (\req(SsR)) so that 
\be
{\sf Q}_L (u) {\sf Q}_R (v) = {\sf Q}_L (v) {\sf Q}_R (u) , \ \ u, v \in \CZ ,
\ele(QLQR)
then define 
\be
Q (v) = {\sf Q}_R (v) {\sf Q}_R (v_0)^{-1} = {\sf Q}_L (v_0)^{-1} {\sf Q}_L (v),
\ele(Qn)
where $v_0$ is  a fixed value  of $v$ so that ${\sf Q}_R (v_0)$ and ${\sf Q}_L (v_0)$ are non-singular. Then $Q(v)$'s form a commuting family and satisfy the $TQ$-relation (\req(8TQ)) ( \cite{B72} (C28) (C37) (C38)). Note that by (\req(Qn)), the operator $Q(v)$, defined up to a normalized factor determined by $v_0$, is independent to the choice of parameters $\tau_{i, j}, \widehat{\tau}_{i, j}$, regardless of the dependence of ${\sf Q}_R, {\sf Q}_L$ on $\tau_{i, j}, \widehat{\tau}_{i, j}$.

\begin{lem} \label{lem:ss} 
The relation $(\req(QLQR))$  is valid for $(\widehat{s}, s)= (2K, 0), (0, 2K)$
in $(\req(LSs))$ , $(\req(SsR))$. 
\end{lem}
{\it Proof}. Define the functions
$$
g(v) := H(v) \Theta (v) (= \frac{1}{\Theta (0)} h(v)), \ ~ \ ~  f(v):= \frac{2 h(v-K)}{h(K)} . 
$$
Using the general formula (\cite{B73II} (C.27)), 
\be
\Theta (2A) H (2B)  - H (2A) \Theta (2B)  = f (A+B) g (A-B),
\ele(fg1) 
one finds the following identity for  $\widehat{s}, s \in \CZ$ and $\lambda, \mu = \pm 1$: 
\bea(c)
\langle   \widehat{s} + 2i \eta  + \lambda  (\eta + u) | s + 2 i' \eta + \mu (\eta - v) \rangle \\ 
= f (\frac{\widehat{s} +s}{2}+ (i + i' + \frac{\lambda +\mu}{2}) \eta + \frac{\lambda  u   - \mu  v}{2} ) g (\frac{\widehat{s} -s}{2} +(i-i'+ \frac{\lambda - \mu}{2}) \eta + \frac{\lambda  u + \mu  v}{2}).
\elea(fg)
By using  (\req(fg)), the product of $\widehat{\sf S}_{i,j} (u)$, ${\sf S}_{i',j'}(v)$ in (\req(LSs)), (\req(SsR)) is expressed by 
$$
\begin{array}{lc}
&\widehat{\sf S}_{i,j} (u) {\sf S}_{i',j'} (v) = \widehat{\tau}_{i, j} {\tau}_{i', j'} f(u, v | i, j ; i', j')g(u, v | i, j ; i', j') , \\
{\rm where} & f(u, v | i, j ; i', j') := f (\frac{\widehat{s} +s}{2}+ (i + i' + \frac{j-i +j'-i'}{2}) \eta + \frac{(j-i)  u   - (j'-i') v}{2} ) , \\
& g(u, v | i, j ; i', j') := g ( \frac{\widehat{s} -s}{2}+ (i - i') \eta + \frac{(j-i-j' +i' )) \eta }{2}+ \frac{ (j-i)  u +(j' - i' ) v }{2} ) .
\end{array}
$$
The relation (\req(QLQR)) will hold if there exist auxiliary functions $P(v, u | n), p(v, u | n)$ for $n \in \ZZ_N$ such that 
\be
\widehat{\sf S}_{i,j} (u) {\sf S}_{i',j'} (v) = p(v, u | i'+i) P(v, u |i'-i) \widehat{\sf S}_{i,j} (v) {\sf S}_{i',j'} (u) P(v, u |j'-j)^{-1} p(v, u | j'+j)^{-1},
\ele(SSpP)
since the product $\widehat{\sf S}_{i,j} (u) {\sf S}_{i',j'} (v)$ differs only by a diagonal gauge transformation when interchanging $v$ and $u$. The condition (\req(SSpP)) 
is equivalent to the following relations:
$$
\begin{array}{ll}
g(u, v | i, j ; i', j') = P(v, u | i'-i) g(v, u | i, j ; i', j') P(v, u | j'-j)^{-1} ,& j-i = -(j'-i')= \pm 1;  \\
f(u, v | i, j ; i', j') = p(v, u | i'+i) f(u, v | i, j ; i', j') p(v, u | j'+j)^{-1} , & j-i = j'-i' = \pm 1 .
\end{array}
$$
The conditions on $g(u, v | i, j ; i', j')$ yield just one condition for $P$: 
$$
\frac{P(v, u | n+2)}{P(v, u | n)} = \frac{g(\frac{\widehat{s} -s}{2}-(n+1) \eta + \frac{u-v}{2}) }{g(\frac{\widehat{s} -s}{2}-(n+1) \eta + \frac{v-u}{2})} \ ~ \ ~ \ ~ ~ {\rm for} \ n \in \ZZ_N.
$$
Since $N$ is odd, the constraint $P(v, u | n+2N)= P(v, u | n)$ in turn yields $\widehat{s} - s = \pm 2K$. Similarly, the conditions on $f(u, v | i, j ; i', j')$ yields the condition on $p$:
$$
\frac{p(v, u | n) }{p(v, u | n+2 )} = \frac{f (\frac{\widehat{s} +s}{2}+ (n + 1) \eta + \frac{ u - v}{2} )}{f (\frac{\widehat{s} +s}{2}+ (n + 1) \eta + \frac{  v   -  u}{2} )}\ ~ \ ~ \ ~ ~ {\rm for} \ n \in \ZZ_N ,  
$$
with $\widehat{s} + s = \pm 2K$. Then follows the values of $\widehat{s}, s$: $(\widehat{s}, s) = ( \pm 2K, 0), (0, \pm 2K)$. Since $\widehat{\sf S }_{i, j}, {\sf S }_{i, j}$ are the same when adding $\pm 4K$ on $\widehat{s}, s$-values, only $(2K, 0), (0, 2K)$ remains as solutions of $(\widehat{s}, s)$ in our discussion. 
$\Box$ \par  \noindent
{\bf Remark}. In the above proof, we indeed show the values of $\widehat{s}, s$ in Lemma \ref{lem:ss} are the only solutions such that the relation (\req(SSpP)) holds.

By Lemma \ref{lem:ss}, there are only two sets of $s , \widehat{s}$-values in the discussion of
$Q$-operator (\req(Qn)) of the root-of-unity eight-vertex model. Moreover these two $Q$-operators can be converted from one to another by the substitution of variables: $v \mapsto v - 2K$. So we need only to consider the $Q$-operator (\req(Qn)) with ${\sf Q}_R$, ${\sf Q}_L$ using the parameters
$$
s = 0 , \ \widehat{s} = 2K . 
$$
Hence the ${\sf S}_{i',j'}, \widehat{\sf S}_{i,j}$ in (\req(SsR)), (\req(LSs)) are zeros except $ j - i = \pm 1$, in which cases\footnote{Note that the free parameter of the $Q_R$-operator in this paper specifies at a value different from that in \cite{F06} Section 3.  In this work, the values of $s, \widehat{s}$ for the $Q_R, Q_L$-operators are distinct, unlike that in \cite{F06} (31) where the free parameter $t$ for $Q_R, Q_L$ takes the same value as in  \cite{Bax} (9.8.38)-(9.8.42) (also see the argument in \cite{F06} Section 4). Indeed, the equality (31) in \cite{F06} (with the same $t$-value) provides only a sufficient condition for the commutativity of the constructed $Q$-operator, which is still valid when the relation (\req(QLQR)) in this paper holds for two (not necessary equal) $s, \widehat{s}$-values.}
\bea(ll)
{\sf S}_{i,j} (v) = 
 |2 i \eta + (j - i ) (\eta - v) \rangle \tau_{i, j} , & 
 \widehat{\sf S}_{i,j} (v) = 
 \widehat{\tau}_{i, j} \langle  2K + 2i \eta  + (j-i)  (\eta + v)|.
\elea(LSR)
The ${\sf S}$-operator of the $\CZ^N$-auxiliary and $\CZ^2$-quantum space is now with the form  
$$
{\sf S} = ({\sf S}_{i, j})_{i, j \in \ZZ_N} = \left( \begin{array}{cccccc}
        0  & {\sf S}_{0,1} & 0&\cdots   & 0&{\sf S}_{0,N-1}\\
        {\sf S}_{1,0} &  0& {\sf S}_{1,2} & \ddots & &\vdots \\
0&&\ddots&\ddots&& \\
\vdots&&\ddots&\ddots&&0 \\
0&&&\ddots& &{\sf S}_{N-2,N-1}\\
{\sf S}_{N-1,0} &0&\cdots &0&{\sf S}_{N-1,N-2}&0
\end{array} \right) ,
$$
similarly for $\widehat{\sf S}$. 
By (\req(HT)), the $Q$-operator satisfies
\be
Q(v - 2K) = S Q(v) = Q(v) S ,
\ele(QS)
where $S$ is the operator in (\req(TRS)).

\subsection{Mathematical verification of functional relations in the eight-vertex model for $\eta = \frac{2 m K}{N}$ \label{ssec.pfQQ}}
Hereafter we set the parameter $(s, \widehat{s}) = (0, 2K)$ in ${\sf Q}_R, {\sf Q}_L$ and  local vectors in (\req(Phi)); so ${\sf S}_{i',j'}, \widehat{\sf S}_{i,j}$ are given by (\req(LSR)).  In this subsection, we show the following theorem about functional relations of the root-of-unity eight-vertex model with $\eta$ in (\req(eta)) by using results in the eight-vertex SOS model \cite{B73I, B73II, B73III, DJKMOp, DJKMO}. 
\begin{thm}\label{thm:Qfun} 
The $Q$-operator $(\req(Qn))$ defined by $(\req(LSR))$ satisfies the $Q$-functional equation $(\req(Qeq))$, equivalent to the $QQ$-relations, $(\req(QQN))$ or  $(\req(QQ))$, with $C$ in $(\req(Cdef))$.
\end{thm}

Associated to the local vector $\Phi_{l, l+ \mu }$ in (\req(Phi)) with $s=0$, we introduce the local covector $\widehat{\Phi}_{l, l+ \mu }$ for $l \in \ZZ , \mu = \pm 1$:
\bea(cl)
\Phi_{l, l+ \mu } (= \Phi_{l, l+ \mu } (v)) = | 2l \eta + \mu (\eta - v) \rangle , &  \\
\widehat{\Phi}_{l, l+ \mu } ( = \widehat{\Phi}_{l, l+ \mu } (v)) = r_{l, l+ \mu} \langle  2K + 2l \eta  + \mu (\eta + v)|, & r_{l, l+ \mu} = \frac{\mu \Theta(0) h(K)}{2h(2\eta) h(2(l+ \mu) \eta-K)}.
\elea(Phi0)
By (\req(fg)), one finds
$$
\bigg( \begin{array}{l}\widehat{\Phi}_{l-1, l }  \\
\widehat{\Phi}_{l+1, l} \end{array} \bigg)(2K+v) \cdot (\Phi_{l, l+1 }, \Phi_{l, l-1 }) (v) = \frac{h(v-\eta)}{h(2\eta) } \bigg( \begin{array}{ll} 1 & 0 \\
0 &  1 \end{array} \bigg) .
$$
For a set of integers $l_1, \ldots, l_{L+1}$ with  $l_{\ell+1} - l_{\ell} = \pm 1$ for $ 1 \leq \ell \leq L$, the product-vector in (\req(psi)) (for $s=0$), and the product-covector are expressed by  
\bea(l)
\psi (l_1, \ldots, l_{L+1})(v) = \Phi_{l_1, l_2}(v) \otimes \Phi_{l_2, l_3}(v) \otimes \cdots \otimes \Phi_{l_L, l_{L+1} }(v), \\
 \widehat{\psi} ( l_1, \ldots, l_{L+1})(v) = \widehat{\Phi}_{l_1, l_2}(v) \otimes \widehat{\Phi}_{l_2, l_3}(v) \otimes \cdots \otimes \widehat{\Phi}_{l_L, l_{L+1}}(v).
\elea(psi0)
The condition (\req(eta)) guarantees the local vector and covector in (\req(Phi0)) unchanged when replacing $l$ by $l \pm N$. Hence for $i, j \in \ZZ_N$ and $j-i = \pm 1$, we shall also write  
$$
\Phi_{i, j} := \Phi_{l, l+(j-i)} (= \Phi_{l'+i-j , l'}) , \ ~ \widehat{\Phi}_{i,j} := \widehat{\Phi}_{l, l+(j-i)} (= \widehat{\Phi}_{l'+i-j, l'}) 
$$
where $l \equiv i$ (or $l' \equiv j)  \pmod{N}$ if no confusion will arise. Similarly, one can write $\psi (i_1, \ldots, i_{L+1})(v) = \psi (l_1, \ldots, l_{L+1})(v)$, $\widehat{\psi} ( i_1, \ldots, i_{L+1} )(v) = \widehat{\psi} ( l_1, \ldots, l_{L+1} )(v)$ for $i_\ell \in \ZZ_N $ and $i_{\ell+1} - i_\ell = \pm 1$ ~ $(1 \leq \ell \leq L)$ using an integer-representative $l_1$ (or $l_{L+1}$) of $i_1$ ($i_{L+1}$ resp.)  in $\ZZ_N$.  
It is expected that product-vectors (covectors) in (\req(psi0)) for all $l_\ell$'s and $v$ span the $2^L$-dimensional vector space $\stackrel{L}{\otimes} \CZ^2$ (the dual space $\stackrel{L}{\otimes} \CZ^{* 2}$ resp.). Equivalently, product-vectors $\psi (l_1, \ldots, l_{L+1})(v_0)$ for $l_1 \in \ZZ_N$ and $l_{\ell+1}- l_\ell = \pm 1$, span the vector space $\stackrel{L}{\otimes} \CZ^2$ for a generic $v_0$; the same for $\stackrel{L}{\otimes} \CZ^{* 2}$ spanned by $\widehat{\psi} ( l_1, \ldots, l_{L+1})(v_0)$'s. We shall use $v_0$ to denote a general value of $v$. For simple notations, we shall also write these vectors at $v_0$ in $\stackrel{L}{\otimes} \CZ^2$ and $\stackrel{L}{\otimes} \CZ^{* 2}$ by 
\be
v (l_1, \ldots, l_{L+1}) := \psi ( l_1, \ldots, l_{L+1})(v_0) , \ ~ h (l_1, \ldots, l_{L+1}) := \widehat{\psi} ( l_1, \ldots, l_{L+1})(v_0) , \ l_1 \in \ZZ_N , 
\ele(epv)
where $l_2, \ldots, l_{L+1}$ are integers with  $l_{\ell+1} - l_{\ell} = \pm 1$ for $ 1 \leq \ell \leq L$.  Note that the number of vectors $v (l_1, \ldots, l_{L+1})$'s is strictly greater than $2^N$. For a general $v_0$, we consider the following linear transformation of $\stackrel{L}{\otimes} \CZ^2$ defined by the sum of products of vectors in (\req(epv)),
\be 
M(v_0) := \sum_{l_1 \in \ZZ_N} {\sum_{l_\ell \in \ZZ}}^\prime v (l_1,  \ldots , l_{L+1}) h (l_1,  \ldots , l_{L+1}) .
\ele(Qv0) 
Hereafter the ''prime'' summation means $l_{\ell+1} - l_\ell = \pm 1$ for $1 \leq \ell \leq L$, and $l_1 \equiv l_{L+1} \pmod{N}$. The operator $M(v_0)$ 
is expected to have the rank $2^N$, hence to be a non-singular operator. Unfortunately I know of no simple way to prove this condition except checking cases by direct computations (some of which will be given in the appendix). Nevertheless we shall assume the non-singular property of the  $\stackrel{L}{\otimes} \CZ^2$-operator (\req(Qv0)) for the rest of this paper.

We now choose some convenient parameter $\tau_{i, j}$, $\widehat{\tau}_{i, j}$ in (\req(LSR)) to represent the $Q$-operator. Denote by ${\sf Q}^0_R$, ${\sf Q}^0_L$ the operators in (\req(QRL)) using the following $\tau_{i, j}$ and $\widehat{\tau}_{i, j}$:
$$
\tau_{i, j} = \widehat{\Phi}_{i, j}(v_0) , \ \ \ \ \widehat{\tau}_{i, j} = r_{i, j} \Phi_{l_1, l_2}(v_0)  ~ \ ~ (j-i= \pm 1) 
$$
where $r_{i, j}$ is defined in (\req(Phi0)). This means the non-zero ${\sf S}^0_{i,j}$ and $\widehat{\sf S}^0_{i,j}$ in (\req(LSR)) are 
\bea(lll)
{\sf S}^0_{i,j} (v) = 
 \Phi_{i, j} (v) \widehat{\Phi}_{i, j}(v_0) , & 
 \widehat{\sf S}^0_{i,j} (v) = \Phi_{l_1, l_2}(v_0)   \widehat{\Phi}_{i,j}(v) ,   & ( j- i= \pm 1).
\elea(LSR0)
In particular, the $Q$ operator in Theorem \ref{thm:Qfun}  can be expressed by
$$
Q (v) = {\sf Q}^0_R (v) {\sf Q}^0_R (v_0)^{-1} = {\sf Q}^0_L (v_0)^{ -1} {\sf Q}^0_L (v).
$$
In order to prove Theorem \ref{thm:Qfun}, we need only to verify the relation (\req(QQN)), which will follows from the equality
\be
T^{(N)}(v- 2\eta ){\sf Q}^0_R(v_0) =  {\sf Q}^0_L (v - 2K) {\sf Q}^0_R (v)  , \ \ v \in \CZ ,
\ele(QTN)
with  $M_0= {\sf Q}^0_L (v_0)^{-1}$ in (\req(QQN)).

\begin{lem} \label{lem:TNPh} 
Let ${\Phi}_{i, j}(v), \widehat{\Phi}_{i',j' }(v)$ be the local vectors in $(\req(Phi0))$, and 
$W^{(N)}(m, m' | l, l')(v- 2 \eta)$ the Nth fusion weights at $v-2 \eta$ in $(\req(WN))$. Then for $m, l \in \ZZ$, $i, i' \in \ZZ_N$ such that $\lambda^{(N)} (:= m-l) = N-1- 2k$ for some $0 \leq k \leq N-1$, and $m \equiv i , l \equiv i' \pmod{N}$, the equality 
$$
W^{(N)}(m, m + \lambda | l, l + \mu )(v- 2 \eta) = \widehat{\Phi}_{i,i+ \lambda }(v-2K) {\Phi}_{i', i'+ \mu}(v)
$$
holds for $\lambda , \mu = \pm 1$. 
\end{lem}
{\it Proof}. 
For $i, i' \in \ZZ_N$ and $\lambda, \mu = \pm 1$, by (\req(fg)), one writes the product $\widehat{\Phi}_{i,j}(v-2K) {\Phi}_{i',j'}(v)$ in the form
$$
\begin{array}{ll}
\widehat{\Phi}_{i,i+ \lambda }(v-2K) {\Phi}_{i', i'+ \mu}(v) = r_{i, i+\lambda} \langle   2i \eta  + \lambda  (\eta + v )  |  2 i' \eta + \mu (\eta - v) \rangle = r_{i, i+\lambda} w_{i ; i-i' }(\lambda, \mu ; v) , 
\end{array}
$$
where  the functions $w_{i ; n }(\lambda, \mu ; v)$ for $i, n \in \ZZ_N$, $\lambda, \mu = \pm 1$ are defined by 
$$
w_{i ; n }(\lambda, \mu ; v) =  \frac{2}{h(K)\Theta (0) } h( -K + (2i - n + \frac{\lambda +\mu}{2}) \eta + \frac{\lambda  - \mu }{2}v) h ( (n+ \frac{\lambda - \mu}{2}) \eta + \frac{\lambda   + \mu  }{2} v). 
$$
Comparing the values in (\req(WN)) with $\widehat{\Phi}_{i,j}(v-2K) {\Phi}_{i',j'}(v)$ with $j=i+ \lambda$, $j'= i'+ \mu$ , we find 
$$
\begin{array}{ll}
W^{(N)}(l-1, l | l'-1, l')(v- 2 \eta ) &= r_{l-1, l} w_{l-1, \lambda^{(N)}}(1,1 ; v) = \widehat{\Phi}_{i,i+1}(v-2K) {\Phi}_{i',i'+1}(v), \\
&(l-1 \equiv i ,  l'-1 \equiv i' ; \lambda^{(N)} = \mu^{(N)} ,  \alpha=    \beta = 1) , 
\\
W^{(N)}(l+1, l | l'+1 , l')(v- 2 \eta)&= r_{l+1, l} w_{l+1, \lambda^{(N)}}(-1,-1 ; v) = \widehat{\Phi}_{i,i-1}(v-2K) {\Phi}_{i',i'-1}(v), \\
&(l+1 \equiv i, l'+1 \equiv i' ; \lambda^{(N)} = \mu^{(N)}, \alpha= \beta = -1 ) , \\
W^{(N)}(l+1, l | l'-1, l')(v-2 \eta ) & = r_{l+1, l}  w_{l+1, \lambda^{(N)}}(-1,1 ; v)
= \widehat{\Phi}_{i,i-1}(v-2K) {\Phi}_{i',i'+1}(v), \\
&( l+1 \equiv i, l'-1 \equiv i' ; \lambda^{(N)} = \mu^{(N)}+2 , \alpha=   - \beta = -1)  \\
W^{(N)}(l-1, l | l'+1, l')(v- 2 \eta) & 
= r_{l-1, l} w_{l-1 ; \lambda^{(N)} }(1, -1 ; v) 
= \widehat{\Phi}_{i,i+1}(v-2K) {\Phi}_{i',i'-1}(v) ,  \\
&(l-1 \equiv i , l'+1 \equiv i' ;
\lambda^{(N)} = \mu^{(N)}-2 , \alpha=   - \beta = 1 ) .
\end{array}
$$
Then follow the results.
$\Box$ \par  \noindent

We now show the relation (\req(QTN)). Using (\req(LSR0)), one can express ${\sf Q}^0_R (v)$ and ${\sf Q}^0_L (v )$ in terms of product-vectors and covectors in (\req(psi0)) and (\req(epv)) :
\bea(ll)
{\sf Q}^0_R (v) &= \sum_{i_\ell \in \ZZ_N, i_1=i_{L+1}} ({\sf S}^0_{i_1,i_2}{\sf S}^0_{i_2,i_3} \cdots {\sf S}^0_{i_L,i_{L+1}})(v) \\
&= \sum_{l_1 \in \ZZ_N} \sum_{l_\ell \in \ZZ} ^\prime \psi (l_1,l_2, \ldots ,l_{L+1})(v) h (l_1,l_2, \ldots ,l_{L+1}) ; \\
{\sf Q}^0_L (v ) & = \sum_{i_\ell \in \ZZ_N, i_1=i_{L+1}} (\widehat{\sf S}^0_{i_1,i_2}\widehat{\sf S}^0_{i_2,i_3} \cdots \widehat{\sf S}^0_{i_L,i_{L+1}})(v) \\
&= \sum_{m_1 \in \ZZ_N} \sum_{m_\ell \in \ZZ} ^\prime v (m_1,m_2, \ldots ,m_{L+1}) \widehat{\psi} (m_1,m_2, \ldots ,m_{L+1})(v) , 
\elea(QLR0)
with the prime summations as before, i.e., $l_{\ell+1} - l_\ell, m_{\ell+1} -m_\ell = \pm 1$ for $1 \leq \ell \leq L$, and $l_1 \equiv l_{L+1}, m_1 \equiv m_{L+1}\pmod{N}$. 
By Lemma \ref{lem:TNPh}, the vectors in (\req(psi0)) has the following product value:
$$
\widehat{\psi} (m_1,m_2, \ldots ,m_{L+1})(v-2K)  \psi (l_1,l_2, \ldots ,l_{L+1})(v) = \prod^L_{\ell=1}W^{(N)}(m_\ell, m_{\ell+1}| l_\ell, l_{\ell+1} )(v- 2 \eta)
$$
when $m_1 - l_1 =  N-1- 2k$ for some $0 \leq k \leq N-1$. Using (\req(QLR0)), one finds 
\bea(rl)
{\sf Q}^0_L (v-2K){\sf Q}^0_R (v)
= \sum_{l_1 \in \ZZ_N} \sum_{m_\ell , l_\ell \in \ZZ} ^\prime & \{ \prod^L_{\ell=1}W^{(N)}(m_\ell, m_{\ell+1}| l_\ell, l_{\ell+1} )(v- 2 \eta) \} \times \\
&  v (m_1, \ldots ,m_{L+1}) h (l_1, \ldots ,l_{L+1}) . 
\elea(Q0)
The vector value of $T^{(N)}(v-2\eta) {\sf Q}^0_R (v_0)$ can be obtained by the relation (\req(TSOS)) for $J=N$:
$$
\begin{array}{c}
\sum_{l_1 \in \ZZ_N} \sum_{l_\ell \in \ZZ} ^\prime T^{(N)}(v-2\eta) \psi (l_1, \ldots ,l_{L+1})(v_0) h (l_1, \ldots ,l_{L+1}) = \\
\sum_{l_1 \in \ZZ_N} \sum_{l_\ell, m_\ell \in \ZZ} ^\prime  \{\prod_{\ell=1}^L W^{(N)}(m_\ell, m_{\ell+1}|l_\ell, l_{\ell+1})(v-2\eta) \} \psi (m_1, \ldots, m_{L+1})(v_0) h (l_1, \ldots ,l_{L+1}) ,
\end{array}
$$
which is the same as (\req(Q0)) by definition of $v (m_1, \ldots ,m_{L+1})$ in (\req(epv)). This in turn yields the equality (\req(QTN)), hence follows Theorem \ref{thm:Qfun}. 
Note that by (\req(epv)), $M^{-1}_0 = {\sf Q}^0_L (v_0 ) (= {\sf Q}^0_R (v_0 ))$ in (\req(QQN)) is equal to the linear transformation $M (v_0)$ in (\req(Qv0)).

Denote $S_R$ the spatial translation operator of $\stackrel{L}{\otimes} \CZ^2$, which takes the $k$th column to $(k+1)$th one for $1 \leq k \leq L$ with the identification $L+1 = 1$. One has 
\bea(ll)
S_R \psi (l_1, l_2, \ldots ,l_{L+1})(v) = \psi (l_0, l_1, \ldots ,l_L)(v), & S_R v (l_1, l_2,  \ldots , l_{L+1}) = v (l_0, l_1, \ldots , l_L) , \\
\widehat{\psi} (l_1, l_2, \ldots , l_{L+1})(v)S_R = \widehat{\psi} (l_2, l_3, \ldots , l_{L+2})(v) , & h (l_1, l_2, \ldots , l_{L+1}) S_R = h (l_2, l_3, \ldots , l_{L+2} ) ,
\elea(SRv)
where $l_0 := l_1 -l_{L+1}+ l_L$, $l_{L+2} := l_{L+1} + l_2-l_1$. By (\req(QLR0)), $S_R$ commutes with  ${\sf Q}^0_R (v )$, ${\sf Q}^0_L (v )$, and $M(v_0)$, hence  
$$
[S_R, Q(v)] = 0 .
$$

\section{Concluding Remarks}\label{sec.F}
By a similar method of producing $Q_{72}$-operator in \cite{B72}, we construct another $Q$-operator, different from $Q_{72}$, of the eight-vertex model at the root-of-unity parameter $\eta$ in (\req(eta)) for the functional-equation study of  the eight-vertex model, as an analogy to functional relations in the superintegrable $N$-state CPM. The $Q$-operator in this work possesses a structure compatible with the Baxter's eight-vertex SOS model in \cite{B73I, B73II, B73III}. By this, using an explicit form of the $Q$-operator and fusion weights of SOS model in \cite{DJKMOp, DJKMO}, we provide a rigorous mathematical argument to show  the functional relations holds under the conjectural non-singular hypothesis (supported by computational evidence in cases) about the operator (\req(Qv0)).  The result could improve our understanding about the newly found ''root-of-unity'' symmetry of the eight-vertex model \cite{De01a, De01b, FM02, FM04, FM410, FM41, FM06}. In this paper we study the symmetry of the eight-vertex model by the method of functional relations, which has been the characteristic features in the theory of CPM (see, e.g. \cite{AMP, BBP} and references therein). Together with results previously known in the superintegrable CPM and the root-of-unity six-vertex models \cite{R04, R05o, R05b, R06Q, R06F}, the conclusion derived from this work has further enhanced the ''universal role'' of CPM among solvable lattice models in regard to the symmetry of  degenerate eigenstates.
Although the analysis is done on the case $\eta$ in (\req(eta)) here, results obtained in \cite{FM02, FM04, FM06} for other  root-of-unity cases strongly suggest that
the analogy can be extended successfully to include all cases in the eight-vertex model with  '' root of unity'' parameter $\eta$. Further progress related to the functional relations of the theory will enrich our knowledge about the symmetry of lattice vertex model, and also serve to demonstrate the universal character of CPM. 

In this paper, we study problems in the root-of-unity eight-vertex model by far involved only with functional relations and the $Q$-operator. However the quantitative nature of the $Q$-operator in our context depends on the choice of $v_0$ in (\req(LSR0)) in performing some  explicit calculations. A convenient $v_0$ has not been found yet, nor the suitable expression of $Q$-eigenvalues as those for $Q_{72}$-operator in  \cite{FM02} (2.22)-(2.25). These works remain to be done. Nevertheless, it is suggested by results obtained in the six-vertex model \cite{DFM,De05, FM00, FM001, R06Q, R06F} that the understanding of symmetry nature of the eight-vertex model should also be related to the study of degenerated eigenstates using the evaluation function in \cite{ De01a, De01b}, and the elliptic current operator recently appeared in \cite{FM06}. 
The connection  with the evaluation function is expected to be found because the better understanding of the symmetry properties depends on it. A process along this line is now under consideration, and further progress would be expected.

\section*{Acknowledgements}
The author is pleased to thank Professor T. Mabuchi for hospitality in November 2006 at the Department of Mathematics, Osaka University, Japan, where part of this work was carried out. 
He also wishes to thank Professor T. Deguchi for useful discussions, and Professor K. Fabricius for helpful communications. This work is supported in part by National Science Council of Taiwan under Grant No NSC 95-2115-M-001-007.

\section*{Appendix: Computation of the $M(v_0)$-operator}
\setcounter{equation}{0}
In this appendix, we provide some computational evidences about the non-singular property of the operator $M(v_0)$ in (\req(Qv0)). By (\req(SRv)), $M(v_0)$ commutes with the spatial translation operator  $S_R$. We shall decompose $M(v_0)$ as the sum of operators of $S_R$-eigenspaces so that one can perform the computations on each factor component for cases with small $L$. For simple notations, we denote 
$$
V = \bigotimes^L \CZ^2 , \ \ V^* = \bigotimes^L \CZ^{2 *} , 
$$
and write the standard basis elements of $V$ and $V^*$ by
$$
|\alpha_1, \cdots, \alpha_L  \rangle := \otimes_\ell | \alpha_\ell \rangle \in ~ V , \ \ \  \langle \alpha_1, \cdots, \alpha_L| := \otimes_\ell \langle \alpha_\ell| \in ~ V^* .
$$
Let $V_k$ be the $S_R$-eigenspace of $V$ with the eigenvalue $ e^{\frac{2 \pi {\rm i} k}{L}}$, hence $V$ has the $S_R$-eigenspace decomposition :
\be
V = \sum_{k=0}^{L-1} V_k . \tag{A1}
\ele(VVk)
For $i \in \ZZ_N$, and $\mu_\ell = \pm 1$ for $1 \leq \ell \leq L$ with $\sum_{\ell =1}^L \mu_\ell \equiv 0 \pmod{N}$, we denote
$$
\begin{array}{l}
\psi [i ; \mu_1, \ldots, \mu_L ] (v) = \psi (l_1, l_2, \cdots, l_{L+1})(v) \in V , \\ \widehat{\psi} [i ; \mu_1, \ldots, \mu_L ] (v) = \widehat{\psi} (l_1, l_2, \cdots, l_{L+1})(v)
 \in V^* ,
 \end{array}
$$
where $l_1 \equiv i \pmod{N}$, and $l_{\ell+1} - l_\ell = \mu_\ell$ for $1 \leq \ell \leq L$. 
As before, $\CZ^N$ is the space of $N$-cyclic vectors $w = \sum_{i \in \ZZ_N} w_i | i \rangle $, and $\CZ^{N *}$ the space of dual vectors $w^* = \sum_{i \in \ZZ_N} w^*_i \langle i |$. We define  
$$
\begin{array}{l}
\psi [\mu_1, \ldots, \mu_L ] (v) := \sum_{i \in\ZZ_N} \psi [i ; \mu_1, \ldots, \mu_L ] (v) \langle i | \in V \otimes \CZ^{N *} , \\
\widehat{\psi} [\mu_1, \ldots, \mu_L ] (v):=  \sum_{i \in\ZZ_N} | i \rangle  \widehat{\psi}  [i ; \mu_1, \ldots, \mu_L ] (v)   \in \CZ^N \otimes  V^* .
 \end{array}
$$
When $v=v_0$, one has $
v (l_1, \ldots, l_{L+1})= \psi ( l_1, \ldots, l_{L+1})(v_0)$ and $h (l_1, \ldots, l_{L+1}) = \widehat{\psi} ( l_1, \ldots, l_{L+1})(v_0)$ by (\req(epv)); and correspondingly, the vector $v(i ; \mu_1, \ldots, \mu_L)$, covector $h (i ; \mu_1, \ldots, \mu_L)$, and $v(\mu_1, \ldots, \mu_L) \in V \otimes \CZ^{N *}$, $h(\mu_1, \ldots, \mu_L) \in \CZ^N \otimes  V^* $.
For $\mu_\ell = \pm 1 ~ (1 \leq \ell \leq L)$ with $\sum_\ell \mu_\ell \equiv 0 \pmod{N}$, we define
$$
\begin{array}{l}
{\sf Q}^0_R [\mu_1, \ldots, \mu_L ] = \sum_{i \in \ZZ_N } \psi [i ; \mu_1, \ldots, \mu_L ] (v) h [i ; \mu_1, \ldots, \mu_L ] =  \psi [\mu_1, \ldots, \mu_L ] (v) h [\mu_1, \ldots, \mu_L ] , \\
{\sf Q}^0_L [\mu_1, \ldots, \mu_L ] = \sum_{i \in \ZZ_N }  v [i ; \mu_1, \ldots, \mu_L ]\widehat{\psi} [i ; \mu_1, \ldots, \mu_L ] (v) =   v [\mu_1, \ldots, \mu_L ]\widehat{\psi}[\mu_1, \ldots, \mu_L ] (v) , \\
\end{array} 
$$
where each second expression in above is given by evaluating covectors of  
$\CZ^{N *}$ on $\CZ^N$-vectors. By (\req(SRv)), one finds
$$
S_R {\sf Q}^0_R [\mu_1, \ldots, \mu_L ](v) S_R^{-1} = {\sf Q}^0_R [\mu_2, \ldots, \mu_{L+1} ](v), \ \
S_R {\sf Q}^0_L [\mu_1, \ldots, \mu_L ](v) S_R^{-1} = {\sf Q}^0_L [\mu_2, \ldots, \mu_{L+1} ](v) ,
$$
where $\mu_{L+1}:= \mu_1$. Consider the group $\langle S_R \rangle$ generated by $S_R$, which acts on basis-vectors $|\mu_1, \ldots, \mu_L \rangle$ with $\sum_\ell \mu_\ell \equiv 0 \pmod{N}$. Denote by ${\cal O}$ the set of $\langle S_R \rangle$-orbits, i.e. the elements ${\sf o} = \langle S_R \rangle |\mu_1, \ldots, \mu_L \rangle$. Define
$$
{\sf Q}^0_{R , {\sf o}}(v) := \sum_{|\mu_1, \ldots, \mu_L \rangle \in {\sf o}} {\sf Q}^0_R [\mu_1, \ldots, \mu_L ](v), \ {\sf Q}^0_{L , {\sf o}}(v) := \sum_{|\mu_1, \ldots, \mu_L \rangle \in {\sf o}} {\sf Q}^0_L [\mu_1, \ldots, \mu_L ](v). 
$$
Then ${\sf Q}^0_{R , {\sf o}}, {\sf Q}^0_{L , {\sf o}}$ commute with $S_R$ and 
$$
{\sf Q}^0_R(v)= \sum_{{\sf o} \in {\cal O}} {\sf Q}^0_{R , {\sf o}}(v), \ {\sf Q}^0_L(v)= \sum_{{\sf o} \in {\cal O}} {\sf Q}^0_{L , {\sf o}}(v). 
$$
We now express  ${\sf Q}^0_{R(v), {\sf o}}$ in terms of the vector-decomposition in (\req(VVk)). Assume ${\sf o}= \langle S_R \rangle | \mu_1, \ldots, \mu_L \rangle$, and let $L_{\sf o}$ be the positive divisor of $L$ such that $\langle S_R^{L_{\sf o}} \rangle$  consists of all elements in $\langle S_R \rangle$ which fix  $| \mu_1, \ldots, \mu_L \rangle$. Then $\mu_{\ell+L_{\sf o}} = \mu_\ell$ for all $\ell$, hence $\sum_{k=0}^{L_{\sf o}-1} \mu_k = \frac{L_{\sf o}}{L} \sum_{k=0}^{L-1} \mu_k \equiv 0 \pmod{N}$, which implies $S_R^{L_{\sf o}}\psi [i ; \mu_1, \ldots, \mu_L ] (v)= \psi [i ; \mu_1, \ldots, \mu_L ] (v)$.   Denote  by $\psi [i ; \mu_1, \ldots, \mu_L ]_k (v)$ the $V_k$-component of $\psi [i ; \mu_1, \ldots, \mu_L ] (v)$ in  (\req(VVk)), and define 
$$
\psi [\mu_1, \ldots, \mu_L ]_k (v) = \sum_{i \in \ZZ_N} \psi [i ; \mu_1, \ldots, \mu_L ]_k (v) \langle i | \in V \otimes \CZ^{N *}.
$$  
Then $\psi [i ; \mu_1, \ldots, \mu_L ]_k (v)=0$ except  $k=$ a multiple of $\frac{L}{L_{\sf o}}$, therefore  
$$
\psi [i ; \mu_1, \ldots, \mu_L ] (v) = \sum_{k = 0}^{L_{\sf o}-1} \psi [i ; \mu_1, \ldots, \mu_L ]_{\frac{kL}{L_{\sf o}}} (v). 
$$
The same statement holds when replacing $\psi$ by $\widehat{\psi}$ . Hence
$$
\begin{array}{cc}
\widehat{\psi} [\mu_1, \ldots, \mu_L ] (v) = \sum^{ L_{\sf o}-1 }_{k = 0} \widehat{\psi} [\mu_1, \ldots, \mu_L ]_{\frac{kL}{L_{\sf o}}} (v),&\psi [\mu_1, \ldots, \mu_L ] (v) = \sum_{k = 0}^{L_{\sf o}-1} \psi [\mu_1, \ldots, \mu_L ]_{\frac{kL}{L_{\sf o}}} (v) , \\
v [\mu_1, \ldots, \mu_L ] = \sum_{k = 0}^{L_{\sf o}-1} v [\mu_1, \ldots, \mu_L ]_{\frac{kL}{L_{\sf o}}}, &
h [\mu_1, \ldots, \mu_L ] = \sum_{k = 0}^{L_{\sf o}-1} h [\mu_1, \ldots, \mu_L ]_{\frac{kL}{L_{\sf o}}}.
\end{array}
$$ 
By 
$$
\begin{array}{l}
{\sf Q}^0_{R , {\sf o}}(v) = \sum_{m=0}^{L_{\sf o}-1} S_R^m {\sf Q}^0_R [\mu_1, \ldots, \mu_L ](v)S_R^{-m}, \\
S_R^m {\sf Q}^0_R [\mu_1, \ldots, \mu_L ](v)S_R^{-m} = \sum_{k, k' = 0}^{L_{\sf o}-1} e^{\frac{2 \pi {\rm i}m (k-k')}{L_{\sf o}}}  \psi [\mu_1, \ldots, \mu_L ]_{\frac{kL}{L_{\sf o}}} (v) h [\mu_1, \ldots, \mu_L ]_{\frac{k'L}{L_{\sf o}}},  
\end{array}
$$
one finds 
$$
{\sf Q}^0_{R , {\sf o}}(v) 
= L_{\sf o} \sum_{k = 0}^{L_{\sf o}-1}   \psi [\mu_1, \ldots, \mu_L ]_{\frac{kL}{L_{\sf o}}} (v) h [\mu_1, \ldots, \mu_L ]_{\frac{kL}{L_{\sf o}}}.
$$
Similarly, $
{\sf Q}^0_{L , {\sf o}}(v) 
= L_{\sf o} \sum_{k = 0}^{L_{\sf o}-1}   v[\mu_1, \ldots, \mu_L ]_{\frac{kL}{L_{\sf o}}} \widehat{\psi} [\mu_1, \ldots, \mu_L ]_{\frac{kL}{L_{\sf o}}}(v)$. Set $v= v_0$ in the above ${\sf Q}^0_R(v)$, then follows
\be
M (v_0 ) = \sum_{{\sf o} \in {\cal O}} M(v_0)_{\sf o}, ~ \ ~  
M(v_0)_{\sf o}  
= L_{\sf o} \sum_{k = 0}^{L_{\sf o}-1}   v[\mu_1, \ldots, \mu_L ]_{\frac{kL}{L_{\sf o}}} h [\mu_1, \ldots, \mu_L ]_{\frac{kL}{L_{\sf o}}} . \tag{A2}
\ele(MSR) 
As the vector and covector in (\req(|v)) are related by $
| v \rangle^t =  \langle -v | \sigma^x$, one finds 
$$
r_{i ; \mu_1, \ldots, \mu_L} \psi [i; \mu_1, \ldots, \mu_L ](-v)^t = \widehat{\psi} [i; \mu_1, \ldots, \mu_L ] (v) R ,
$$
where $R$ is the spin-reflection operator in (\req(TRS)), and $r_{i ; \mu_1, \ldots, \mu_L}= \prod_{\ell=1}^L r_{l_\ell, l_{\ell +1}} $ with $l_\ell = i + \sum_{k=1}^{\ell -1} \mu_k$. Hence 
$$
r_{\mu_1, \ldots, \mu_L} \psi [\mu_1, \ldots, \mu_L ](-v)^t = \widehat{\psi} [\mu_1, \ldots, \mu_L ] (v) R 
$$
where $r_{\mu_1, \ldots, \mu_L} = \sum_{i \in \ZZ_N} r_{i ; \mu_1, \ldots, \mu_L} | i \rangle \langle i |$ the non-degenerate diagonal matrix of $\CZ^N$. Note that $r_{\mu_1, \ldots, \mu_L}= r_{\mu_2, \ldots, \mu_{L+1}}$, which depends only the $\langle S_R \rangle$-orbit ${\sf o}$ of $|\mu_1, \ldots, \mu_L \rangle$. We shall also write $r_{\sf o}  = r_{\mu_1, \ldots, \mu_L}$. Then $M(v_0)_{\sf o} $ is expressed by
\be
M(v_0)_{\sf o}  
= L_{\sf o} \sum_{k = 0}^{L_{\sf o}-1}   \psi [\mu_1, \ldots, \mu_L ]_{\frac{kL}{L_{\sf o}}} (v_0) r_{\sf o} \psi [\mu_1, \ldots, \mu_L ]_{\frac{kL}{L_{\sf o}}} (-v_0)^t R . \tag{A3}
\ele(prp)
Since $R$ commutes with $S_R$, the above summation provides the decomposition of $M(v_0)_{\sf o}$ on subspaces $V_{\frac{kL}{L_{\sf o}}}$.

We now use the formulas (\req(MSR)),(\req(prp)) to determine the non-singular property of $M(v_0)$ for $L=2,4$, where $\sum_\ell \mu_\ell \equiv 0 \pmod{N}$ is equivalent to $\sum_\ell \mu_\ell = 0$. For convenience, we denote
$$
H_i (v) : = H ((2i+1) \eta +v) , \ ~ \ ~ \Theta_i (v) : = \Theta ((2i+1) \eta +v) , \ \ \ i \in \ZZ_N .
$$ 
Then $H_i (-v) = - H_{-i-1} (v)$ and $\Theta_i (-v) = \Theta_{-i-1} (v)$.
For $L=2$,  ${\cal O}$ consists of one $\langle S_R \rangle$-orbit ${\sf o}=\{ |1,-1 \rangle, |-1,1 \rangle  \}$ with $L_{\sf o}=2$, and the decomposition  $V = V_0 + V_1$ of (\req(VVk)) is given by 
$$
V_0 = \CZ |1,1 \rangle + \CZ |-1,-1 \rangle + \CZ (|1,-1 \rangle +|-1,1 \rangle), \ ~ \ ~ V_1 = \CZ (|1,-1 \rangle - |-1,1 \rangle ).
$$
With $(\mu_1, \mu_2) = (1, -1)$, one has $\psi [i; 1, -1](v) = \psi (i+1, i, i+1)(v)$, equal to  $H_i(-v)H_i(v) |1,1 \rangle + H_i(-v) \Theta_i(v) |1,-1 \rangle + \Theta_i(-v) H_i(v) |-1,1 \rangle  + \Theta_i(-v) \Theta_i(v) |-1,-1 \rangle$. Using the equalities (by (\req(fg1))) 
$$
\begin{array}{l}
H_i(-v) \Theta_i(v)+\Theta_i (-v)H_i(v) 
= \frac{2  h((2i+1) \eta) h(v +K)}{\Theta (0) h(K)} , \\ 
\Theta_i (v) H_i (- v) -H_i(v) \Theta_i(-v) = \frac{2 h((2i+1)\eta-K) h(v )}{\Theta (0) h(K)} ,
\end{array}
$$
we find $\psi [i; 1, -1] = \psi [i; 1, -1]_0 + \psi [i; 1, -1]_1$ with 
$$
\begin{array}{rl}
\psi [i; 1, -1]_0(v)=& H_i(-v)H_i(v) |1,1 \rangle + \Theta_i(-v) \Theta_i(v) |-1,-1 \rangle + \frac{  h((2i+1) \eta) h(v +K)}{\Theta (0) h(K)} (|1,-1 \rangle + |-1,1 \rangle) , \\
\psi [i; 1, -1]_1(v) = & \frac{ h((2i+1)\eta-K) h(v )}{\Theta (0) h(K)}  (|1,-1 \rangle - |-1,1 \rangle) .
\end{array}
$$
Hence $\psi [1, -1](v) = \psi [1, -1]_0(v) + \psi [1, -1]_1(v)$ where $\psi [1, -1]_k(v) =  \sum_{i \in \ZZ_N} \psi [i; 1, -1]_k (v)\langle i |$ for $k=0, 1$. By (\req(MSR)) and (\req(prp)), 
$$
M(v_0) R = 2 \sum_{k=0, 1}\psi [1, -1 ]_k (v_0) r  \psi [1, -1 ]_k (-v_0)^t = 2 \sum_{k=0, 1}(-1)^k \psi [1, -1 ]_k (v_0) r  \psi [1, -1 ]_k (v_0)^t ,  
$$
where $r = \sum_{i \in \ZZ_N} r_{i, i+1} r_{i+1, i} | i \rangle \langle i |$.  The non-singular $M(v_0)$ is equivalent to the non-degenerate bilinear form  $\psi [1, -1 ]_k (v_0) r  \psi [1, -1 ]_k (v_0)^t$ of $V_k$ for $k=0,1$, which is obvious for $k=1$. For $k=0$, $\psi [1, -1]_0(v_0)$ defines three linear independent $N$-cyclic vectors, then by the non-zero entries of $r$, follows the non-singular $M(v_0)_{|V_0}$. 

We now consider the case $L=4$, where the decomposition $V = \sum_{k=0}^3 V_k $ in (\req(VVk)) is given by
\bea(ll)
V_0  =  \CZ {\sf v} + \CZ {\sf v}' + \CZ {\sf x}_0  + \CZ {\sf y}_0 + \CZ {\sf w}_0 + \CZ {\sf u}_0 , &
V_1 =   \CZ {\sf x}_1  + \CZ {\sf y}_1 + \CZ {\sf u}_1 , \\
V_2 =  \CZ {\sf x}_2  + \CZ {\sf y}_2 + \CZ {\sf w}_1 + \CZ {\sf u}_2 , &
V_3 =   \CZ {\sf x}_3  + \CZ {\sf y}_3 + \CZ {\sf u}_3, \tag{A4}
\elea(V4)
where 
$$
\begin{array}{ll}
{\sf v} = | 1,1,1,1 \rangle, & {\sf v}' = |-1,-1,-1,-1 \rangle, \\ 
{\sf x}_n = \sum_{k=0}^3 {\rm i }^{nk} S_R^k |1,1,1, -1 \rangle , &
{\sf y}_n = \sum_{k=0}^3 {\rm i }^{nk} S_R^k  |1,-1,-1, -1 \rangle, \\
{\sf w}_m = \sum_{k=0}^1 (-1)^{mk} S_R^k |1,-1,1, -1 \rangle ,&
{\sf u}_n = \sum_{k=0}^3 {\rm i }^{nk} S_R^k  |1,1,-1, -1 \rangle ,
\end{array}
$$
for $0 \leq n \leq 3$, $m=0,1$. The transport of the above basis elements, denoted by ${\sf v}^*, {\sf v}^{' *}$, ${\sf x}^*_n, {\sf y}^*_n$, ${\sf w}^*_m$ and ${\sf u}^*_n$, form a basis of $V^*$. The set ${\cal O}$ in (\req(MSR)) consists of two $\langle S_R \rangle$-orbits:
$$
\begin{array}{lll}
{\sf o}_1 = &\{|1,-1, 1, -1 \rangle , |-1,1,-1,1  \rangle \} ,& L_{{\sf o}_1}=2 ;  \\
{\sf o}_2 = &\{ |1,1, -1, -1 \rangle, |-1,1, 1, -1 \rangle, |-1,-1, 1, 1 \rangle, |1,-1, -1, 1 \rangle \} , & L_{{\sf o}_2 }=4 .
\end{array}
$$     
For the orbit ${\sf o}_1$ represented by $|1,-1, 1, -1 \rangle$, one has $\psi [i;1,-1, 1, -1](v) = \psi ( i, i+1, i, i+1, i)(v)$ ( = $\psi (i, i+1, i)(v) \otimes \psi (i, i+1, i)(v)$), which is equal to  
$$
\begin{array}{l}
H_i(v)^2 H_i(-v)^2 {\sf v}+  \Theta_i(v)^2 \Theta_i(-v)^2 {\sf v}'  
\\+H_i(-v)^2H_i(v)\Theta_i(v) (|1,1,1,-1 \rangle + |1,-1,1,1 \rangle) 
+H_i(-v)H_i(v)^2\Theta_i(-v) (|-1,1,1,1 \rangle + |1,1,-1,1 \rangle) \\
+H_i(-v) \Theta_i(-v)   \Theta_i(v)^2 (|1,-1,-1,-1 \rangle +|-1,-1,1,-1 \rangle) \\
+ H_i(v) \Theta_i(-v)^2 \Theta_i(v)  (|-1,1,-1,-1 \rangle +|-1,-1,-1,1 \rangle) 
\\
+ H_i(-v)^2 \Theta_i(v)^2 |1,-1,1,-1 \rangle  + H_i(v)^2  \Theta_i(-v)^2 |-1,1,-1,1 \rangle  
+  H_i(-v)H_i(v)  \Theta_i(-v) \Theta_i(v){\sf u}_0 .
\end{array}
$$
In the basis in (\req(V4)), one finds $\psi [i;1,-1, 1, -1]= \psi [i;1,-1, 1, -1]_0 + \psi [i;1,-1, 1, -1]_2$ where 
\bea(ll)
\psi [i;1,-1, 1, -1]_0(v)=&
H_i(v)^2 H_i(-v)^2 {\sf v}+  \Theta_i(v)^2 \Theta_i(-v)^2 {\sf v}'  
\\
&+\frac{H_i(-v)^2H_i(v)\Theta_i(v) +H_i(-v)H_i(v)^2\Theta_i(-v)}{2} {\sf x}_0 \\
&+\frac{H_i(-v) \Theta_i(-v)   \Theta_i(v)^2 + H_i(v) \Theta_i(-v)^2 \Theta_i(v)}{2}{\sf y}_0  
\\
&+ \frac{H_i(-v)^2 \Theta_i(v)^2 + H_i(v)^2  \Theta_i(-v)^2}{2} {\sf w}_0 +   H_i(-v)H_i(v)  \Theta_i(-v) \Theta_i(v){\sf u}_0 ;  \\
\psi [i;1,-1, 1, -1]_2 (v) =&\frac{H_i(-v)^2H_i(v)\Theta_i(v)  -H_i(-v)H_i(v)^2\Theta_i(-v)}{2}{\sf x}_2 \\
&+\frac{H_i(-v) \Theta_i(-v)   \Theta_i(v)^2 -H_i(v) \Theta_i(-v)^2 \Theta_i(v)}{2} {\sf y}_2
\\
&+\frac{H_i(-v)^2 \Theta_i(v)^2 - H_i(v)^2  \Theta_i(-v)^2}{2}   {\sf w}_1 . \tag{A5}
\elea(p41)
Hence $\psi [1,-1, 1, -1] = \psi [1,-1, 1, -1]_0 + \psi [1,-1, 1, -1]_2$, with 
$$
\psi [1,-1, 1, -1]_k(v) = \sum_{i \in \ZZ_N} \psi [i;1,-1, 1, -1]_k (v) \langle i |, \ \ k=0, 2.
$$ 
By (\req(prp)), one obtains
\bea(ll)
M(v_0)_{{\sf o}_1} R &
= 2 \sum_{k = 0}^1   \psi [1,-1, 1, -1]_{2k } (v_0) r_{{\sf o}_1} \psi [1,-1, 1, -1]_{2k}(-v_0)^t \\
&= 2 \sum_{k = 0}^1 (-1)^k  \psi [1,-1, 1, -1]_{2k } (v_0) r_{{\sf o}_1} \psi [1,-1, 1, -1]_{2k}(v_0)^t , \tag{A6}
\elea(M41)
where $r_{{\sf o}_1} = \sum_{i \in \ZZ_N} r_{i, i+1}^2 r_{i+1, i}^2|i \rangle \langle i |$. 

For ${\sf o}_2$ (= the class of $|1,1,-1,-1 \rangle$), $\psi [i;1,1,-1, -1](v) = \psi ( i, i+1, i+2, i+1, i)(v)$  is expressed by  
$$
\begin{array}{l}
H_i (-v)H_{i+1}(-v) H_i (v)  H_{i+1}(v) {\sf v} +
\Theta_i (-v) \Theta_{i+1}(-v) \Theta_i (v) \Theta_{i+1}(v) {\sf v}' +  \\
H_i (-v) H_{i+1}(-v)H_{i+1}(v) \Theta_i (v)|1,1,1,-1 \rangle + 
H_{i+1}(-v) H_i (v) H_{i+1}(v)\Theta_i (-v) |-1,1,1,1 \rangle + \\
H_i (-v)H_i (v)H_{i+1}(v) \Theta_{i+1}(-v) |1,-1,1,1 \rangle + H_i (-v)H_{i+1}(-v) H_i (v)\Theta_{i+1}(v) |1,1,-1,1 \rangle + \\
H_i (-v) \Theta_{i+1}(-v) \Theta_i (v) \Theta_{i+1}(v) |1,-1,-1,-1 \rangle + H_{i+1}(-v) \Theta_i (-v)\Theta_i (v)  \Theta_{i+1}(v) |-1,1,-1,-1 \rangle + \\
H_{i+1}(v) \Theta_i (-v) \Theta_{i+1}(-v)\Theta_i (v) |-1,-1,1,-1 \rangle + 
H_i (v) \Theta_i (-v) \Theta_{i+1}(-v) \Theta_{i+1}(v) |-1,-1,-1,1 \rangle +  \\
H_i (-v)H_{i+1}(v)  \Theta_{i+1}(-v) \Theta_i (v)  |1,-1,1,-1 \rangle + H_i (v)   H_{i+1}(-v) \Theta_i (-v) \Theta_{i+1}(v) |-1,1,-1,1 \rangle +\\
H_i (-v)H_{i+1}(-v) \Theta_i (v)   \Theta_{i+1}(v)  |1,1,-1,-1 \rangle + H_{i+1}(-v)H_{i+1}(v) \Theta_i (-v)\Theta_i (v)   |-1,1,1,-1 \rangle + \\
H_i (v) H_{i+1}(v) \Theta_i (-v) \Theta_{i+1}(-v)  |-1,-1,1,1 \rangle +H_i (-v)H_i (v)   \Theta_{i+1}(-v) \Theta_{i+1}(v)  |1,-1,-1,1 \rangle. 
\end{array}
$$
In the basis in (\req(V4)), $\psi [i;1,1, -1, -1]= \sum_{k=0}^3 \psi [i;1,1,-1,-1]_k$ where 
\bea(c)
\psi [ i; 1,1,-1,-1]_0(v) = \\
H_i (-v)H_{i+1}(-v) H_i (v)  H_{i+1}(v) {\sf v} +
\Theta_i (-v) \Theta_{i+1}(-v) \Theta_i (v) \Theta_{i+1}(v) {\sf v}'  \\
+\bigg(H_i (-v) H_{i+1}(-v)H_{i+1}(v) \Theta_i (v) + 
H_{i+1}(-v) H_i (v) H_{i+1}(v)\Theta_i (-v)   \\
+H_i (-v)H_i (v)H_{i+1}(v) \Theta_{i+1}(-v)   + H_i (-v)H_{i+1}(-v) H_i (v)\Theta_{i+1}(v)\bigg) \frac{{\sf x}_0 }{4} \\
+\bigg(H_i (-v) \Theta_{i+1}(-v) \Theta_i (v) \Theta_{i+1}(v)  + H_{i+1}(-v) \Theta_i (-v)\Theta_i (v)  \Theta_{i+1}(v)   \\
+H_{i+1}(v) \Theta_i (-v) \Theta_{i+1}(-v)\Theta_i (v)  + 
H_i (v) \Theta_i (-v) \Theta_{i+1}(-v) \Theta_{i+1}(v)\bigg)\frac{{\sf y}_0}{4} +  \\
+\bigg(H_i (-v)H_{i+1}(v)  \Theta_{i+1}(-v) \Theta_i (v) + 
H_i (v)   H_{i+1}(-v) \Theta_i (-v) \Theta_{i+1}(v)\bigg)\frac{{\sf w}_0}{4}   \\
+\bigg(H_i (-v)H_{i+1}(-v) \Theta_i (v)   \Theta_{i+1}(v)  + H_{i+1}(-v)H_{i+1}(v) \Theta_i (-v)\Theta_i (v)  \\
+H_i (v) H_{i+1}(v) \Theta_i (-v) \Theta_{i+1}(-v)  +H_i (-v)H_i (v)   \Theta_{i+1}(-v) \Theta_{i+1}(v) \bigg) \frac{{\sf u}_0}{4}  , \tag{A7}
\elea(p420)
\bea(c)
\psi [ i; 1,1,-1,-1]_1(v) = \\
\bigg(H_i (-v) H_{i+1}(-v)H_{i+1}(v) \Theta_i (v) - 
{\rm i} H_{i+1}(-v) H_i (v) H_{i+1}(v)\Theta_i (-v)    \\
+(-{\rm i})^2 H_i (-v)H_i (v)H_{i+1}(v) \Theta_{i+1}(-v)   + (-{\rm i})^3 H_i (-v)H_{i+1}(-v) H_i (v)\Theta_{i+1}(v) \bigg)\frac{{\sf x}_1}{4}  \\
+\bigg(H_i (-v) \Theta_{i+1}(-v) \Theta_i (v) \Theta_{i+1}(v)  - {\rm i} H_{i+1}(-v) \Theta_i (-v)\Theta_i (v)  \Theta_{i+1}(v)   \\
+(-{\rm i})^2 H_{i+1}(v) \Theta_i (-v) \Theta_{i+1}(-v)\Theta_i (v) + 
(-{\rm i})^3 H_i (v) \Theta_i (-v) \Theta_{i+1}(-v) \Theta_{i+1}(v) \bigg) \frac{{\sf y}_1}{4}   \\
+\bigg( H_i (-v)H_{i+1}(-v) \Theta_i (v)   \Theta_{i+1}(v) - {\rm i} H_{i+1}(-v)H_{i+1}(v) \Theta_i (-v)\Theta_i (v))  \\
+(-{\rm i})^2 H_i (v) H_{i+1}(v) \Theta_i (-v) \Theta_{i+1}(-v) +(-{\rm i})^3 H_i (-v)H_i (v)   \Theta_{i+1}(-v) \Theta_{i+1}(v)\bigg) \frac{{\sf u}_1}{4}, \tag{A8}
\elea(p421)
\bea(c)
\psi [ i; 1,1,-1,-1]_2(v) = \\
\bigg(H_i (-v) H_{i+1}(-v)H_{i+1}(v) \Theta_i (v)-
H_{i+1}(-v) H_i (v) H_{i+1}(v)\Theta_i (-v)   \\
+H_i (-v)H_i (v)H_{i+1}(v) \Theta_{i+1}(-v)  - H_i (-v)H_{i+1}(-v) H_i (v)\Theta_{i+1}(v)\bigg) \frac{{\sf x}_2}{4}  \\
+\bigg(H_i (-v) \Theta_{i+1}(-v) \Theta_i (v) \Theta_{i+1}(v) - H_{i+1}(-v) \Theta_i (-v)\Theta_i (v)  \Theta_{i+1}(v)   \\
+H_{i+1}(v) \Theta_i (-v) \Theta_{i+1}(-v)\Theta_i (v) - 
H_i (v) \Theta_i (-v) \Theta_{i+1}(-v) \Theta_{i+1}(v)\bigg) \frac{{\sf y}_2}{4}   \\
+\bigg(H_i (-v)H_{i+1}(-v) \Theta_i (v)   \Theta_{i+1}(v) - H_{i+1}(-v)H_{i+1}(v) \Theta_i (-v)\Theta_i (v)  \\
+H_i (v) H_{i+1}(v) \Theta_i (-v) \Theta_{i+1}(-v)  - H_i (-v)H_i (v)   \Theta_{i+1}(-v) \Theta_{i+1}(v)\bigg) \frac{{\sf u}_2}{4}  \\
+ \bigg(H_i (-v)H_{i+1}(v)  \Theta_{i+1}(-v) \Theta_i (v)  -
H_i (v)   H_{i+1}(-v) \Theta_i (-v) \Theta_{i+1}(v)\bigg)\frac{{\sf w}_1}{2}, \tag{A9}
\elea(p422)
\bea(c)
\psi [ i; 1,1,-1,-1]_3(v) = \\
\bigg(H_i (-v) H_{i+1}(-v)H_{i+1}(v) \Theta_i (v) + 
{\rm i} (H_{i+1}(-v) H_i (v) H_{i+1}(v)\Theta_i (-v))    \\
+{\rm i}^2(H_i (-v)H_i (v)H_{i+1}(v) \Theta_{i+1}(-v) )   + {\rm i}^3 (H_i (-v)H_{i+1}(-v) H_i (v)\Theta_{i+1}(v)) \bigg)\frac{{\sf x}_3}{4} \\
+\bigg(H_i (-v) \Theta_{i+1}(-v) \Theta_i (v) \Theta_{i+1}(v)  + {\rm i}( H_{i+1}(-v) \Theta_i (-v)\Theta_i (v)  \Theta_{i+1}(v) ) + \\
{\rm i}^2 (H_{i+1}(v) \Theta_i (-v) \Theta_{i+1}(-v)\Theta_i (v) ) + 
{\rm i}^3 (H_i (v) \Theta_i (-v) \Theta_{i+1}(-v) \Theta_{i+1}(v)) \bigg) \frac{{\sf y}_3}{4}  \\ + \bigg( H_i (-v)H_{i+1}(-v) \Theta_i (v)   \Theta_{i+1}(v) + {\rm i}( H_{i+1}(-v)H_{i+1}(v) \Theta_i (-v)\Theta_i (v)) )+ \\
{\rm i}^2( H_i (v) H_{i+1}(v) \Theta_i (-v) \Theta_{i+1}(-v))  +{\rm i}^3 (H_i (-v)H_i (v)   \Theta_{i+1}(-v) \Theta_{i+1}(v))\bigg) \frac{{\sf u}_3}{4}. \tag{A10}
\elea(p423)
Hence $\psi [1,1,-1, -1](v) = \sum_{k=0}^3 \psi [1,1,-1, -1]_k (v)$ with 
$$
\psi [1,1, -1, -1]_k(v) = \sum_{i \in \ZZ_N} \psi [i;1,1, -1, -1]_k (v) \langle i |, \ \ 0 \leq k \leq 3.
$$ 
By (\req(prp)), 
\bea(ll)
M(v_0)_{{\sf o}_2} R &
= 4 \sum_{k = 0}^3   \psi [1,1,-1, -1]_k (v_0) r_{{\sf o}_2} \psi [1,1,-1,-1]_k(-v_0)^t \tag{A11}
\elea(M422)
where $r_{{\sf o}_2} = \sum_{i \in \ZZ_N} r_{i, i+1}r_{i+1, i+2} r_{i+2, i+1} r_{i+1, i}|i \rangle \langle i |$. By (\req(M41)) and (\req(M422)), $M(v_0) R ~ (= M(v_0)_{{\sf o}_1} R + M(v_0)_{{\sf o}_2} R) $ is expressed by 
\bea(c)
2 \psi [1,-1, 1, -1]_0 (v_0) r_{{\sf o}_1} \psi [1,-1,1,-1]_0 (v_0)^t 
+  4\psi [1,1,-1,-1]_0 (v_0) r_{{\sf o}_2} \psi [1,1,-1,-1]_0 (v_0)^t  \\
- 2   \psi [1,-1, 1, -1]_2 (v_0) r_{{\sf o}_1} \psi [1,-1, 1, -1]_2(v_0)^t 
+ 4\psi [1,1,-1, -1]_2 (v_0) r_{{\sf o}_2} \psi [1,1,-1,-1]_2(-v_0)^t 
\\ + 4\psi [1,1,-1,-1]_1 (v_0) r_{{\sf o}_2} \psi [1,1,-1,-1]_1(-v_0)^t  
+ 4\psi [1,1,-1,-1]_3 (v_0) r_{{\sf o}_2} \psi [1,1,-1,-1]_3(-v_0)^t , \tag{A12}
\elea(M4)
which induces the $V_k$-endomorphism $M(v_0)_{|V_k} R$  for $0 \leq k \leq 3$. The non-singular property of $M(v_0)$ will follow from the non-degeneracy of $M(v_0)_{|V_k} R$  for all $k$. For $k=1, 3$, $V_k$ is 3-dimensional with the basis in (\req(V4)), by which
the three $N$-cyclic vectors to express  $\psi [1,1,-1,-1]_k (v_0)$ in (\req(p421)) and (\req(p423)) are linear independent for a generic $v_0$; and the same for $\psi [1,1,-1,-1]_k(-v_0)$. Therefore $M(v_0)_{|V_k}$ is a non-singular automorphism of $V_k$. For $k=0$, by (\req(p41)) and (\req(p420)), there are six $N$-cyclic vectors in each of $\psi [1,-1,1,-1]_0 (v_0)$ and $\psi [1,1,-1,-1]_0 (v_0)$ for a generic $v_0$. Then by using different quadratic forms of $\CZ^N$, they form the $V_0$-automorphism $M(v_0)_{|V_0}R$. The non-zero determinant of $M(v_0)_{|V_0}R$ is expected by direct computation for a given $N$. Also for $k=2$, one can argue the non-singular property of $M(v_0)_{|V_2}$ in a similar manner. For example in the case $N=3$, using
$H_i (-v) = - H_{-i-1} (v)$ and $\Theta_i (-v) = \Theta_{-i-1} (v)$, one can express (\req(p41)),(\req(p420)) and (\req(p422)) in terms of $H_i (:= H_i(v)), \Theta_i (:= \Theta_i(v))$ for $i=0,1,2$, then obtains 
$$
\begin{array}{ll}
\psi [1,-1, 1, -1]_0(v) &= 
{\sf v} \otimes (H_0^2 H_2^2 , H_1^4, H_0^2H_2^2) +  {\sf v}' \otimes(\Theta_0^2 \Theta_2^2, \Theta_1^4 , \Theta_0^2 \Theta_2^2) \\
&+ \frac{(H_2^2H_0\Theta_0 -H_2H_0^2\Theta_2){\sf x}_0}{2}  \otimes (1,0,-1) +
\frac{(-H_2 \Theta_2 \Theta_0^2 + H_0 \Theta_2^2 \Theta_0){\sf y}_0}{2} \otimes(1,0,-1)\\
& + 
 \frac{{\sf w}_0}{2}  \otimes( H_0^2  \Theta_2^2+ H_2^2 \Theta_0^2 , 2H_1^2 \Theta_1^2, H_0^2 \Theta_2^2 + H_2^2  \Theta_0^2) \\&-
{\sf u}_0 \otimes (H_0 H_2 \Theta_0\Theta_2, H_1^2 \Theta_1^2, H_0H_2 \Theta_0 \Theta_2); \\

\psi [1,1,-1,-1]_0(v) &= 
(H_0H_2){\sf v} \otimes (H_1^2, H_1^2, H_0 H_2) +  (\Theta_0\Theta_2){\sf v}' \otimes(\Theta_1^2 ,\Theta_1^2 , \Theta_0 \Theta_2) \\
&+ \frac{(H_2  \Theta_0 -H_0 \Theta_2)H_1^2{\sf x}_0}{4}  \otimes ( 1,
- 1, 0)+
\frac{( H_2\Theta_0  - H_0  \Theta_2) 
 \Theta_1^2{\sf y}_0}{4} \otimes(-1,  1, 0  ) \\
&- \frac{{\sf w}_0}{4}  \otimes(H_1\Theta_1(H_2 \Theta_0 + H_0 \Theta_2 ),H_1 \Theta_1(H_2  \Theta_0  +    H_0  \Theta_2),  H_0^2  \Theta_2^2 + 
H_2^2 \Theta_0^2 )
\\
&+\frac{(H_0H_1 \Theta_1 \Theta_2+ H_0 H_2\Theta_1^2 
+H_1 H_2 \Theta_0 \Theta_1 -H_1^2 \Theta_0 \Theta_2 ){\sf u}_0}{4} \otimes (1,
1, 0); \\
\psi [1,-1, 1, -1]_2(v) &= 
\frac{{\sf x}_2}{2} \otimes (H_0 H_2^2 \Theta_0 +H_0^2H_2\Theta_2, 2H_1^3\Theta_1 , H_0 H_2^2 \Theta_0 +H_0^2H_2\Theta_2)\\ 
&- \frac{{\sf y}_2}{2} \otimes (H_2  \Theta_0^2 \Theta_2 +H_0\Theta_0\Theta_2^2 , 2H_1\Theta_1^3 , H_2  \Theta_0^2 \Theta_2 +H_0\Theta_0\Theta_2^2 )  \\
&+ \frac{{\sf w}_1}{2}  \otimes (H_2^2 \Theta_0^2 - H_0^2  \Theta_2^2, 
-2H_1^2 \Theta_1^2 , H_0^2 \Theta_2^2 - H_2^2  \Theta_0^2); \\

\psi [1,1,-1,-1]_2(v) &= 
\frac{{\sf x}_2}{4} \otimes (H_1^2 H_2 \Theta_1 +H_0H_1^2 \Theta_2  -2H_0H_1H_2 \Theta_1 , \\
&H_0H_1H_2 \Theta_2 +H_0 H_1H_2\Theta_1  
-H_1^2H_2 \Theta_0  - H_0H_1^2\Theta_2,H_0^2H_2\Theta_0   - H_0 H_2^2\Theta_0)\\
&+\frac{(H_0 \Theta_1^2\Theta_2   
+H_2  \Theta_0\Theta_1^2 - 2H_1 \Theta_0\Theta_1 \Theta_2){\sf y}_2}{4} \otimes (-1, 1 , 0) \\
&+ \frac{{\sf u}_2}{4} \otimes (H_1H_2 \Theta_0\Theta_1 + H_1^2 \Theta_0\Theta_2 
+H_0H_1 \Theta_1\Theta_2   + H_0H_2  \Theta_1^2 , \\
&H_1H_2 \Theta_0\Theta_1 + H_1^2 \Theta_0\Theta_2 
+H_0H_1 \Theta_1\Theta_2   + H_0H_2  \Theta_1^2 ,
4H_0 H_2  \Theta_0 \Theta_2   ) \\
&+ \frac{{\sf w}_1}{2} \otimes (-H_1H_2\Theta_0 \Theta_1 +
H_0H_1\Theta_1 \Theta_2 , 
-H_1 H_2 \Theta_0\Theta_1  + H_0H_1\Theta_1 \Theta_2,
-H_0^2  \Theta_2^2 +
H_2^2 \Theta_0^2).
\end{array}
$$ 
Using (\req(M4)), one finds the non-singular $M(v_0)_{|V_k} R$ when $v_0$ is generic for $k=0,2$.
For a general $N$, the zero-determinant of (\req(M4)) would provide a complicated relation among theta functions $H_i(v)$'s and $\Theta_j(v)$'s, which is unlikely to be true. Unfortunately we can not provide a rigorous mathematical argument about this statement.

\end{document}